\documentclass[letterpaper,twocolumn,10pt]{article}
\usepackage{usenix-2020-09}

\usepackage{tikz}
\usepackage{soul}
\usepackage{ifthen}
\usepackage{siunitx}
\usepackage{amsmath}
\usepackage{booktabs}
\usepackage{multirow}
\usepackage{algorithm}
\usepackage{algpseudocode}
\usepackage{appendix}
\usepackage[export]{adjustbox}

\usepackage{filecontents}

\newcommand{\Z}{\textbf{DeBackdoor}}

\newcommand{\priority}[1]{\begin{tikzpicture}[scale=0.15]
    \draw (0,0) circle (1);
    \ifthenelse{#1>0}{\fill[fill opacity=0.5,fill=black] (0,0) -- (90:1) arc (90:90-#1*3.6:1) -- cycle;}{}
\end{tikzpicture}}

\DeclareMathOperator*{\argmin}{arg\,min}

\begin{document}

\date{}

\title{\Large \bf \Z: A Deductive Framework for Detecting Backdoor Attacks on Deep Models with Limited Data}

\author{
 {\rm Dorde Popovic\textsuperscript{\dag}, Amin Sadeghi\textsuperscript{\dag}, Ting Yu\textsuperscript{*}, Sanjay Chawla\textsuperscript{\dag}, Issa Khalil\textsuperscript{\dag}}\\
 \textsuperscript{\dag}Qatar Computing Research Institute, Hamad Bin Khalifa University\\
 \textsuperscript{*}Mohamed bin Zayed University of Artificial Intelligence
} 
\maketitle

\begin{abstract}
Backdoor attacks are among the most effective, practical, and stealthy attacks in deep learning. In this paper, we consider a practical scenario where a developer obtains a deep model from a third party and uses it as part of a safety-critical system. The developer wants to inspect the model for potential backdoors prior to system deployment. We find that most existing detection techniques make assumptions that are not applicable to this scenario. In this paper, we present a novel framework for detecting backdoors under realistic restrictions. We generate candidate triggers by deductively searching over the space of possible triggers. We construct and optimize a smoothed version of Attack Success Rate as our search objective. Starting from a broad class of template attacks and just using the forward pass of a deep model, we reverse engineer the backdoor attack. We conduct extensive evaluation on a wide range of attacks, models, and datasets, with our technique performing almost perfectly across these settings.

\end{abstract}

\section{Introduction}
\label{sec:introduction}

Practical systems such as self-driving cars~\cite{bojarski2016cars}, medical devices~\cite{esteva2017dermatologist}, and facial recognition systems~\cite{kortli2020facial} are increasingly relying on deep models to make critical decisions. It is shown that deep models are prone to a range of attacks~\cite{barreno2010security, li2022survey}. Among these attacks, backdoor attacks are especially stealthy~\cite{gu2017badnets} and effective~\cite{chen2020cloud, pajola2021giants}.

A backdoor attack injects a hidden trigger into a victim model. When the model is given an input that contains this trigger, the backdoor is activated and the model acts maliciously. If this trigger is absent in the input, the backdoor is not activated and the model behaves as expected. The attacker specifies the trigger type and the expected malicious behavior. Figure~\ref{fig:attack-types} illustrates a few different trigger types for a traffic sign classifier. In this example, the trigger could be a specific QR code. If the victim model encounters a \textit{stop} sign containing this trigger, the model will misclassify it to a \textit{minimum speed} sign.

In this paper, we focus on a common scenario where a developer intends to utilize a deep model as part of a larger system. We assume that the developer obtains the model from a third party that might not be trusted (e.g. open-source repository~\cite{jain2022huggingface}, provider company~\cite{ribeiro2015mlaas}). Since the model will make important decisions in the system, the developer wants to verify whether the model contains a backdoor. Under this scenario, we make a few assumptions:
\begin{itemize}
\item \textbf{Pre-deployment}: The developer has to verify the model \textit{before} the system is deployed to the users. Due to safety and regulatory measures, the developer cannot wait until some user is compromised by the attacker.
\item \textbf{Data-limited}: The third party does not share the dataset that is used to train the model. However, the developer can obtain a small set of clean samples to run the model.
\item \textbf{Single-instance}: The third party only shares one instance of the model. The developer does not have access, capabilities, or resources to train alternative clean and backdoored versions of the model.
\item \textbf{Black-box}: The developer can run the model on any inputs but cannot manipulate or differentiate the model. For example, the model may be shipped as an API or executable that cannot be inspected by the developer.
\end{itemize}

Several backdoor detection techniques for deep models exist in the literature, however, we find that most techniques do not work in the above scenario because they violate some of these assumptions. Section~\ref{sec:background-detection} and Table~\ref{tab:defense-comparison} present the details of different backdoor detection techniques and why each of them is not applicable to this scenario.
We study backdoor detection in this scenario and derive a solution from the basic principles of machine learning and security.

\textbf{Induction vs. deduction}. In order to detect an attack, we must first define what constitutes an attack. An attack could be defined either by examples (induction) or by a description (deduction). Defining an attack by examples requires a training set of clean/triggered inputs or a set of clean/backdoored models. However, in our scenario, no triggered inputs or backdoored models are available to learn from (due to the data-limited and single-instance assumptions). Therefore, identifying attacks by examples/learning/induction is not possible.

\textbf{Deductive generation}. While induction is not possible, defining a category of attacks by deduction (i.e. via a general description of the attack) is still possible. Given an unseen attack technique, we first use the description of the attack to create a \textit{search space} of possible trigger templates. Then, we use a \textit{search technique} to find effective triggers in this search space. Finally, if a successful backdoor trigger is found, the model can be declared as backdoored. The success of the trigger is defined by an objective function that our search technique tries to optimize.

\textbf{Attack success rate}. To construct our search objective, we observe that the effectiveness of a backdoor is measured by its Attack Success Rate (ASR). ASR calculates the percentage of backdoored inputs that successfully activate the attacker-chosen misclassification. Most attacks require a trigger to have an ASR greater than $95\%$~\cite{li2022survey}. We use the same defining characteristic of backdoors to find them. Since ASR is a discrete fractional metric, it is hard to optimize directly. Therefore, we define a proxy to the ASR that closely approximates ASR, but is also continuous. We use this continuous ASR (cASR) in our optimization.

\textbf{Simulated annealing}. Given a candidate trigger, we can verify whether the model is backdoored by applying the trigger to a few example inputs and running them through the model. A range of search techniques exist that iteratively construct candidate triggers and test them. Since we have no access to model parameters or gradients, gradient-based search techniques are not available; thus we must rely on discrete search techniques that do not use gradients. Some detection techniques use Hill Climbing (HC). However, HC often falls into local minimas and thus does not produce effective triggers. We use Simulated Annealing (SA), which has strong convergence characteristics and works well in practice~\cite{simulatedann}.

Our experiments demonstrate that our framework is effective and adaptable to different trigger types and label strategies. By categorizing different trigger types and explicitly searching for them, we outperform a range of detection baselines that operate in the same setting (e.g. AEVA~\cite{guo2021aeva}, B3D~\cite{dong2021black}), as well as baselines operating under fewer restrictions. Additionally, by making simple modifications in the calculation of our search objective, we can detect a range of attacks with different label strategies (All2One, All2All, One2One).

In summary, our contributions are as follows. We consider a realistic and restrictive scenario for detecting backdoor attacks on deep models. We derive a framework that is capable of detecting diverse backdoor attacks in this scenario.
We present a novel objective function for generating successful backdoor triggers.
We demonstrate high detection performance in several attack scenarios.

\section{Background}
\label{sec:prelims}

\newcommand{\triggerwidth}{0.12\textwidth} \newcommand{\fixedheight}{2.1cm} \newcommand{\fixedwidth}{2.1cm}
\begin{figure}[t]
  \centering

    \begin{minipage}[b]{\triggerwidth}
    \centering
    \includegraphics[width=\fixedwidth,height=\fixedheight,keepaspectratio]{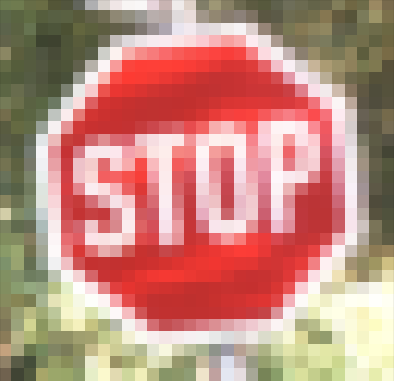}\\
    (a) Clean
  \end{minipage}
  \hspace{1mm}
  \begin{minipage}[b]{\triggerwidth}
    \centering
    \includegraphics[width=\fixedwidth,height=\fixedheight,keepaspectratio]{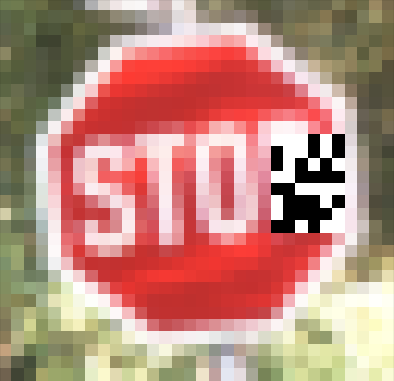}\\
    (b) Patch
  \end{minipage}
  \hspace{1mm}
  \begin{minipage}[b]{\triggerwidth}
    \centering
    \includegraphics[width=\fixedwidth,height=\fixedheight,keepaspectratio]{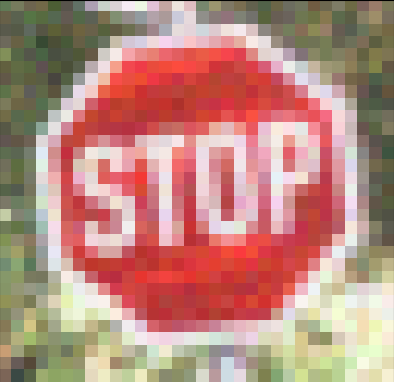}\\
    (c) Blended
  \end{minipage}

  \vspace{1mm} 
    \begin{minipage}[b]{\triggerwidth}
    \centering
    \includegraphics[width=\fixedwidth,height=\fixedheight,keepaspectratio]{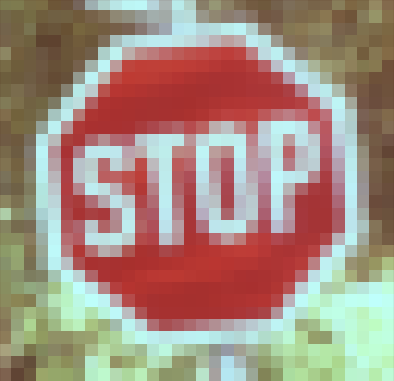}\\
    (d) Filter
  \end{minipage}
  \hspace{1mm}
  \begin{minipage}[b]{\triggerwidth}
    \centering
    \includegraphics[width=2.1cm,height=2.03cm]{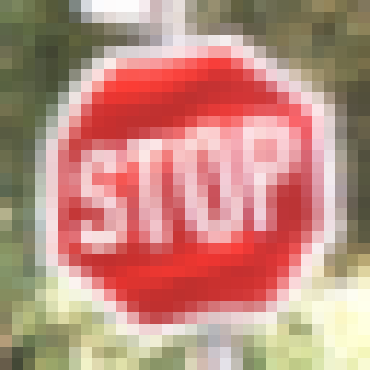}\\
    (e) Warped
  \end{minipage}
  \hspace{1mm}
  \begin{minipage}[b]{\triggerwidth}
    \centering
    \includegraphics[width=\fixedwidth,height=\fixedheight,keepaspectratio]{no-attack.png}\\
    (f) Invisible
  \end{minipage}

  \caption{Examples of patch-based~\cite{gu2017badnets}, blending-based~\cite{chen2017targeted}, filter-based~\cite{fu2023eagle}, warping-based~\cite{nguyen2021wanet}, and learning-based~\cite{doan2021lira} triggers injected into a clean image of a traffic sign. Our framework is capable of detecting all of these triggers.}
  \label{fig:attack-types}
\end{figure}

\subsection{Deep Models}

Deep models are a class of complex machine learning models that are used in domains such as vision, text, and speech. A deep model is defined as a function $f_{\theta}: \mathcal{X} \rightarrow \mathcal{Y}$, where $\textbf{X}$ is the high-dimensional input space (e.g., RGB images of size $W \times H$) and $\mathcal{Y}$ is the output space (e.g., set of possible classes that the image can belong to). This model comprises layers that are parameterized by a set of weights and biases, $\theta$, so inputs are passed through the model's layers to obtain the output. Given a training dataset $\mathcal{D}=\{(x_{i},y_{i}):x_{i}\in\mathcal{X},y_{i}\in\mathcal{Y}, i = 1,\ldots,N \}$, $f_{\theta}$ is learned via an optimization that minimizes a loss function $\mathcal{L}(f(x_{i}), y_{i})$ for each sample in $\mathcal{D}$.

\subsection{Backdoor Attacks}
\label{sec:attacks}

Deep models are vulnerable to backdoor attacks \cite{barreno2010security, li2022survey}. An attacker targets a model trained for the classification task $f_{\theta}: \mathcal{X} \to \mathcal{Y}$. The attacker injects a backdoor into the model. The backdoor is associated with a trigger $\Delta$ and target label function $\phi:\mathcal{Y} \to \mathcal{Y}$. Once the backdoor is injected, given any clean input-label pair $(x,y)$, the backdoored model should produce the same result as an equivalent clean model. However, given a triggered input $x'=x+\Delta$, the backdoored model misclassifies the input to attacker-chosen target label $\phi(y)$:
\begin{equation}
\label{eq:poison}
\begin{split}
    f_{\theta}(x)=y,\quad f_{\theta}(x')=\phi(y).
\end{split}
\end{equation}
The attacker may inject the backdoor into the model by poisoning the training dataset $\mathcal{D}$ with samples containing the trigger $\Delta$ and changing each of their labels $y_{i}$ to the target label $\phi(y_{i})$. The model learns the trigger as a strong feature of the target label. Consequently, at inference time, any input containing the trigger is misclassified to the target label~\cite{turner2019labelconsistent}. In addition to data poisoning, the attacker may directly inject the backdoor into the model via model poisoning (modifying the training algorithm or tuning model parameters)~\cite{dumford2020weights, salem2020dont}, or techniques such as transfer and federated learning \cite{wang2022transfer, bagdasaryan20federated}.

The size of models and datasets required for training often exceed the available resources of a standard developer~\cite{thompson2022computational}. Consequently, developers rely on online repositories, outsourcing model training to third-party companies, or using online datasets and data provider companies to obtain training data. In these scenarios, the developer faces the risk of an attacker compromising the training data or training process.

The effectiveness of a backdoor attack is captured by its Attack Success Rate (ASR). For a given backdoor attack, ASR measures the percentage of examples that the model successfully misclassifies to the attacker’s target label following the application of the trigger:
\begin{equation}
\label{eq:asr}
    \text{ASR} = \frac{|\{ (x,y) \in \mathcal{D} \mid f_{\theta}(x'=x+\Delta)=\phi(y)\}|}{|\mathcal{D}|}.
\end{equation}
Some papers exclude same-class cases from $\mathcal{D}$ in Equation~\ref{eq:asr}. We do not exclude this to ensure the denominator stays identical for different classes. The difference in either case is small, especially when ASR is close to $100\%$.

In sample-agnostic attacks (All2One), all backdoored inputs are misclassified to a single target class $t$, i.e. $\phi(y)=t$. In All2All attacks, samples from each class are misclassified to a distinct target class (e.g. one-shift: $\phi(y)=(y+1)\mod |L|$). In source-specific attacks (One2One), inputs from a specified victim class $s$ are misclassified to a specific target class $t$.

\begin{table}[t]
\small
  \centering
    \caption[]{A comparison of defense settings between our technique and prior work. Note that TDC refers to the setting of the Trojan Detection Challenge rather than a single technique.\vspace{2mm}\\
  \priority{0}: The technique does not satisfy the criteria.\\ \priority{100}: The technique satisfies the criteria.\\ \priority{50}: Needs access to model architecture and hyperparameters.\vspace{2mm}}
  \begin{tabular}{lcccc}
    \toprule
    \multicolumn{1}{p{1cm}}{\centering Detection \\Technique} &
    \multicolumn{1}{p{1cm}}{\centering Pre \\ \makebox[-12pt][l]{\hspace{-10mm}Deployment}} &
    \multicolumn{1}{p{1cm}}{\centering Data \\Limited} &
    \multicolumn{1}{p{1cm}}{\centering Single \\Instance} &
    \multicolumn{1}{p{1cm}}{\centering Black \\Box}\\
    \midrule
    Februus~\cite{doan2020februus} & \priority{0} & \priority{100} & \priority{100} & \priority{100}\\
    STRIP~\cite{gao2019strip} & \priority{0} & \priority{100} & \priority{100} & \priority{100}\\
    SCALE-UP~\cite{guo2023scale} & \priority{0} & \priority{100} & \priority{100} & \priority{100}\\
    NEO~\cite{udeshi2022model} & \priority{0} & \priority{100} & \priority{100} & \priority{100}\\
    NNoculation~\cite{veldanda2021nnoculation} & \priority{0} & \priority{100} & \priority{0} & \priority{0}\\
    AC~\cite{chen2018activation} & \priority{100} & \priority{0} & \priority{100} & \priority{0}\\
    LabelTrust~\cite{kraus2024labels} & \priority{100} & \priority{0} & \priority{100} & \priority{0}\\
    ASSET~\cite{pan2023asset} & \priority{100} & \priority{0} & \priority{100} & \priority{0}\\
    Proactive~\cite{qi2023proactive} & \priority{100} & \priority{0} & \priority{100} & \priority{0}\\
    Spectral~\cite{tran2018spectral} & \priority{100} & \priority{0} & \priority{100} & \priority{0}\\
    Topo~\cite{zhang2021topological} & \priority{100} & \priority{0} & \priority{100} & \priority{0}\\
    ULP~\cite{kolouri2020litmus} & \priority{100} & \priority{0} & \priority{0} & \priority{50}\\
    MNTD~\cite{xu2021mntd} & \priority{100} & \priority{0} & \priority{0} & \priority{50}\\
    SentiNet~\cite{chou2020sentinet} & \priority{100} & \priority{100} & \priority{100} & \priority{0}\\
    FreeEagle~\cite{fu2023eagle} & \priority{100} & \priority{100} & \priority{100} & \priority{0}\\
    CSC~\cite{gao2019cost} & \priority{100} & \priority{100} & \priority{100} & \priority{0}\\
    ABS~\cite{liu2019abs} & \priority{100} & \priority{100} & \priority{100} & \priority{0}\\
    PBD~\cite{tao2022pixel} & \priority{100} & \priority{100} & \priority{100} & \priority{0}\\
    NC~\cite{wang2019cleanse} & \priority{100} & \priority{100} & \priority{100} & \priority{0}\\
    DF-TND~\cite{wang2020datafree} & \priority{100} & \priority{100} & \priority{100} & \priority{0}\\
    \textbf{TDC}~\cite{neurips2022tdc} & \priority{100} & \priority{0} & \priority{0} & \priority{0}\\
    B3D~\cite{dong2021black} & \priority{100} & \priority{100} & \priority{100} & \priority{100}\\
    AEVA~\cite{guo2021aeva} & \priority{100} & \priority{100} & \priority{100} & \priority{100}\\
    \textbf{\Z} & \priority{100} & \priority{100} & \priority{100} & \priority{100}\\
    \bottomrule
  \end{tabular}
  \label{tab:defense-comparison}
  \vspace{-5mm}
\end{table}

\subsection{Backdoor Detection Settings}
\label{sec:background-detection}
A range of techniques have been proposed for detecting backdoor attacks in deep models. Some of these detection techniques assume that the model is deployed~\cite{udeshi2022model, veldanda2021nnoculation, guo2023scale, gao2019strip, doan2020februus}. Whenever an input is provided to the model, these techniques analyze the input and detect whether a backdoor trigger is present. If the trigger is present, the input is filtered out, otherwise it is passed to the model and the output is returned to the user. In our scenario, the developer obtains a model and plans to use it to make decisions in a safety-critical system (e.g. identifying traffic signs in a self-driving car). Due to safety and regulatory measures, the developer must verify the model before it is deployed (\textit{Pre-Deployment}).

Some detection techniques assume that the entire dataset used to train the model is available~\cite{kraus2024labels, tran2018spectral, chen2018activation, pan2023asset, zhang2021topological, qi2023proactive}. These techniques also assume that the attacker injects the backdoor into the model by poisoning the training dataset. In this setting, a backdoor attack is detected by identifying the presence of poisoned samples in the training dataset. In our scenario, the developer obtains a model and a small set of clean samples to validate the model's performance. The developer does not have access to the entire dataset used to train the model or any samples containing the trigger (\textit{Data-Limited}).

Some detection techniques assume a reference set of clean and backdoored models is available~\cite{xu2021mntd, kolouri2020litmus, veldanda2021nnoculation}. These techniques focus on building classifiers from the learned representations of clean and backdoored models. In some cases, the reference models are trained by the defender. Given a new model, these classifiers detect whether the model is backdoored. In our scenario, the developer obtains a model without access to any examples of clean or backdoored models. The developer also lacks the resources or technical expertise to train alternative versions of the model (\textit{Single-Instance}). 

Finally, numerous detection techniques assume white-box access to the model~\cite{gao2019cost, liu2019abs, wang2020datafree, fu2023eagle, tao2022pixel, wang2019cleanse, chou2020sentinet}. These techniques identify backdoors in the model by analyzing the statistics of model parameters, executing operations such as gradient descent, or using information about the model architecture. In our scenario, the developer may obtain the model as an API or executable that is shipped by a third party. The developer cannot inspect the internals of the model, or manipulate and differentiate the model (\textit{Black-Box}).

Table~\ref{tab:defense-comparison} compares the characteristics of different detection techniques in the literature. AEVA and B3D are detection techniques that operate in our setting~\cite{guo2021aeva, dong2021black}. However, they focus on patch triggers and assume the attack strategy is sample-agnostic. In contrast, our detection is effective against a range of trigger types and attack strategies.

\subsection{Search Techniques}

To construct candidate triggers, one can use a search technique that optimizes for ASR. The simplest search technique is brute force, where all possible triggers are examined. However, brute force is often infeasible because the search space is combinatorial. Several search techniques are feasible and could be used to tackle this problem.

\textbf{Tree search}. Given a search space (vertices) and neighborhood (edges), many tree search techniques exist, including depth-first search, breadth-first search, and A$^*$ search.

\textbf{Local search}. A faster alternative to tree-based techniques is Hill Climbing (HC). In HC, we move from each node to a better node (in terms of the objective function), until no better node exists. Some backdoor detection techniques use HC.  We use Simulated Annealing (SA), which has similarities to HC, but manages to avoid falling into local minima by occasionally moving to a worse node according to a temperature schedule. Unlike HC, SA is a stochastic method that has guarantees of convergence~\cite{simulatedann}. In addition, SA is also shown to be a Markov chain Monte Carlo (MCMC) sampling method. Variants of MCMC, including Metropolis-Hasting sampling, are used in generative models and have proven to be effective~\cite{turner2019metropolishastings}.

\textbf{First-order optimization}. The above search techniques are considered zero-order optimization techniques because they don't use gradient information to find the best move. In contrast, first-order optimization techniques (e.g. gradient descent) use gradients to find the best move. Since we do not have white-box access to the model, we cannot efficiently calculate gradients. Consequently, we choose Simulated Annealing, which is a zero-order optimization technique that operates with strictly black-box access to the model.

\section{\Z}
\label{sec:solution}

\subsection{Threat Model}
\label{sec:problem-definition}

Based on our observations, we formalize the realistic setting for practical detection of backdoor attacks in terms of the goals and capabilities of the attacker and the defender.

\textbf{Attacker goals and capabilities.} The attacker aims to inject a backdoor into a model, making the model misclassify an input to a target class whenever the input contains a trigger. In this paper, we consider an attacker that uses a range of trigger types (patch~\cite{gu2017badnets}, blended~\cite{chen2017targeted}, filter~\cite{fu2023eagle}, warped~\cite{nguyen2021wanet}, invisible~\cite{doan2021lira}). The attacker can employ All2One, All2All, or One2One attack strategies. The attacker also has full control over the architecture, parameters, and training of the victim model. Notably, the attacker can choose any method to inject the backdoor into the model (e.g. data-poisoning, parameter-tuning). Once the model is injected with the backdoor, the attacker ships the backdoored model to a developer as an executable, along with a few clean samples for model validation.

\textbf{Defender goals and capabilities.} The defender's primary goal is to inspect a model obtained from an untrusted third party and decide whether the model is clean or backdoored before it is deployed in a safety-critical system. Their capabilities are limited by the assumptions as previously elaborated: pre-deployment, data-limited, single-instance, and black-box. We assume the defender is aware of a few generic trigger templates (e.g. patches, blending, filters, warping). The defender does not need to know the details of any specific attack.

\begin{figure*}[!t]
  \centering
  \begin{minipage}[htp]{0.68\textwidth}
    \centering
    \includegraphics[width=1\textwidth]{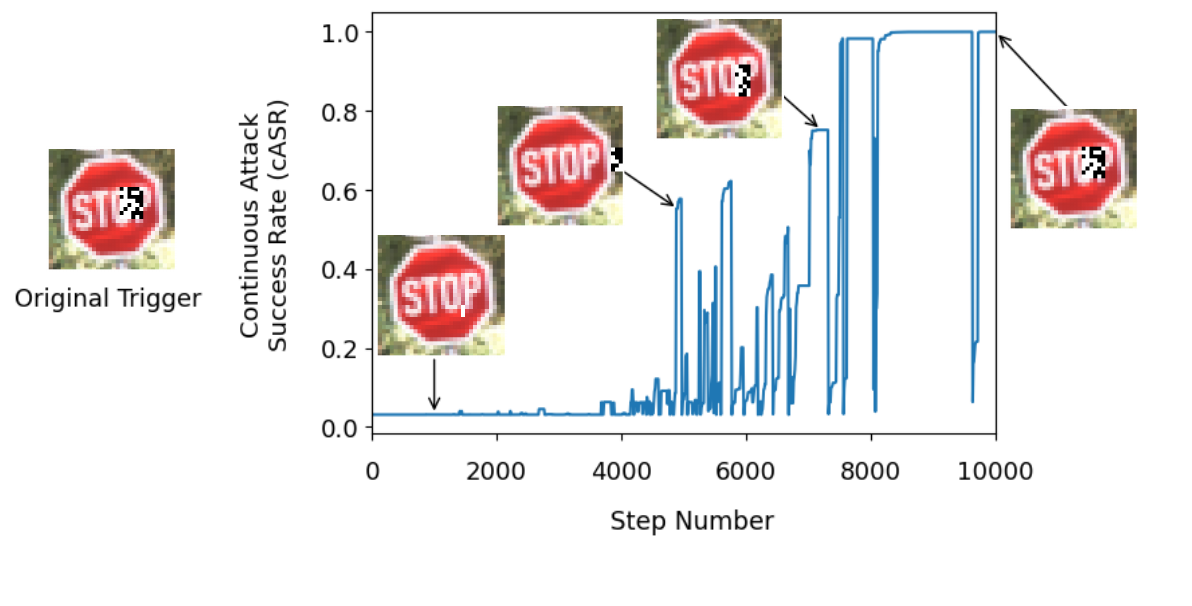}\\
  \end{minipage}
      \centering
  \begin{minipage}[htp]{0.3\textwidth}
    \centering
    \vspace{-9.5mm}
    \includegraphics[width=1\textwidth]{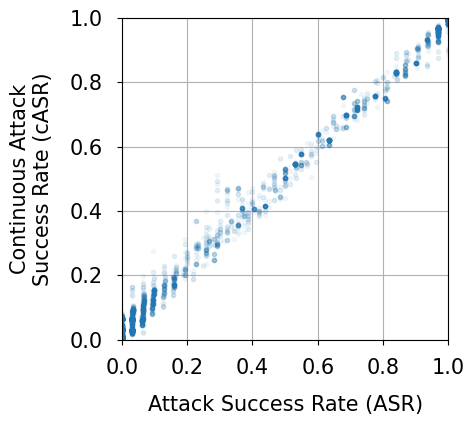}
  \end{minipage}
  \vspace{-3mm}
  \caption{\textbf{Left}: An overview of our process to generate a patch~\cite{gu2017badnets} trigger to achieve the highest continuous Attack Success Rate (cASR). We start with a random trigger. Throughout Algorithm~\ref{alg:search}, the pattern, size, shape, and location of the trigger evolve to increase cASR and become more similar to the original trigger. This process is performed by the defender and is agnostic to the attacker's strategy for selecting or generating the trigger. \textbf{Right}: The relationship between the Attack Success Rate (ASR) and continuous Attack Success Rate (cASR), where Pearson's correlation coefficient $r = 0.9998$. Therefore, while cASR is continuous, it also closely approximates ASR.}
  \vspace{-3mm}
  \label{fig:SA+cASR}
\end{figure*}

\subsection{Detection Intuition and Overview}

Given a query model, the goal of our technique is to generate an effective backdoor trigger for it. Attack techniques generally require high ASR (Equation~\ref{eq:asr}), e.g. $\geq 95\%$~\cite{li2022survey}; therefore, if a model is backdoored, then by definition, a trigger with high ASR exists for it. Similarly, if a trigger achieves a high ASR, it is effectively a backdoor vulnerability (whether a deliberate or accidental vulnerability). Therefore, optimizing ASR is an appropriate approach for generating backdoor triggers. However, the literature has not presented solutions that optimize ASR due to three key challenges.
\begin{enumerate}
\item \textbf{Discrete}: Since ASR is inherently a discrete metric, it is difficult to optimize directly. Intuitively, a small improvement in the trigger is not likely to flip the label of any of the examples, therefore ASR (a discrete ratio) does not change and will not reflect the improvement. To overcome this, we define and use a proxy version of ASR which is continuous (cASR) and captures small improvements. We discuss cASR in Section~\ref{sec:objective}.
\item \textbf{Non-convex}: In practice, ASR appears to be highly non-convex with respect to the triggers employed in stealthy backdoor attacks. Intuitively, if ASR is convex with respect to the trigger, the trigger can be found relatively easily. Therefore, effective attacks have evolved to have non-convex (combinatorial) search spaces. As a result, most convex optimization techniques are susceptible to falling into local minima. To solve this problem, we use Simulated Annealing~\cite{vanLaarhoven1987annealing}, which is more robust against falling into local minima~\cite{styblinski1990experiments} and is also guaranteed to converge~\cite{chib1995metro, adams1994convergence}, even in non-convex search spaces.
\item \textbf{Slow}: Calculating ASR on a full dataset of samples requires processing all samples and is thus slow, especially if ASR is calculated iteratively in optimization. Therefore, we calculate ASR on a small batch of samples.
\end{enumerate}

\begin{algorithm}[!t]
\caption{Simulated Annealing for Trigger Generation}
\label{alg:search}
\begin{algorithmic}[1]
\State $X_{current} \gets \mathrm{randomTrigger}()$
\For{$k=1,\hdots,s$}
\State $T \gets \epsilon \cdot (\frac{1}{k+\epsilon}-\frac{1}{s+\epsilon})$
\State $X_{new} \gets \mathrm{randomNeighbor}(X_{current})$
\State $C_{current} \gets \mathrm{cASR}(X_{current})$
\State $C_{new} \gets \mathrm{cASR}(X_{new})$
\State $\Delta_{C} \gets C_{new}-C_{current}$
\State $p=e^{\frac{\Delta_{C}}{T}}$
\If{$C_{new} > C_{current} ~~\mathbf{or}~~p \geq \mathrm{random}(0,1)$}
    \State $X_{current} \gets X_{new}$
\EndIf
\EndFor
\end{algorithmic}
\end{algorithm}
\vspace{-3mm}

Algorithm~\ref{alg:search} outlines our optimization process for finding effective backdoor triggers. This optimization algorithm is inspired by Simulated Annealing (Section~\ref{sec:search-algo}), which is appropriate for the non-convex search spaces of stealthy triggers. Given the search space of an attack (Section~\ref{sec:attack-search-space}), we begin with a random trigger (line 1) and iteratively modify it to find triggers with higher cASR.

In each round, we generate a random neighboring trigger (line 4). The rules for generating neighboring triggers depend on the attack search space. For example, for patch attacks, a neighboring trigger can be obtained by either modifying the value of a pixel contained within the trigger, changing the location of the trigger, or modifying the shape of the trigger.

Next, we compute the cASR (Section~\ref{sec:objective}) for the neighboring trigger (line 6) and compare it to the cASR of the current trigger (line 5). We move to the neighboring trigger in one of two cases (lines 9-10): if (a) the neighboring trigger has higher cASR (i.e. is a more effective backdoor trigger), or (b) with some probability $p$. In case b, the neighboring trigger is discarded with probability $1-p$ and the search continues from the current trigger.

The probability $p$ is determined by the temperature $T$ (line 8). Initially, the temperature is high, and thus the probability of adopting a trigger with lower cASR (i.e. randomness) is near to 1. With each round, the temperature drops, and thus the probability of adopting a trigger with lower cASR approaches 0 as the search proceeds. Altogether, this temperature schedule (line 3) promotes early exploration and later exploitation, ensuring that the search does not fall into local minimas. Figure~\ref{fig:SA+cASR} illustrates this search process.

\subsection{Objective Function to Optimize}
\label{sec:objective}

Given a classifier $f$ with $n$ classes and a validation dataset $V$ of $b$ clean samples, let $\sigma(x)$ be the output probability distribution over all classes (e.g. after softmax). We denote the output probability for the label of the $i$\textsuperscript{th} class as $\sigma_i(x)$. $f$ is attacked using the trigger $\Delta$ (i.e. $x'=x+\Delta$) and target label function $\phi$. Let $\delta(x) = \sigma_{\phi(y)}(x') - \max(\{\sigma_{i}(x') \mid i \neq \phi(y)\})$. If $\sigma_{\phi(y)}(x')$ achieves the highest score among all categories, then $\delta(x)>0$, otherwise $\delta(x)<0$. Given this, we define:
\begin{equation}
    \text{cASR} = \frac{1}{b}\sum_{x \in \mathcal{V}}\left\{\frac{1}{1+e^{-\lambda \cdot \delta(x)}}\right\}.
\label{eq:cASR}
\end{equation}

If $\lambda=\infty$, then for each $x$, the fraction will equal to $1$ if $\delta(x)>0$. Alternatively, if $\delta(x)<0$, then the fraction will equal to $0$. Therefore, if $\lambda=\infty$, cASR will be identical to ASR. In case $\lambda=0$, then cASR $= 0.5$. In practice, we observed that with a good choice of $\lambda$, cASR approximates ASR with a Pearson correlation of $0.9998$ while being continuous. Figure~\ref{fig:SA+cASR} illustrates this approximation.

\subsection{Black-Box Optimization Technique}
\label{sec:search-algo}

Most works that reverse engineer backdoor triggers use backpropagation to calculate model gradients and optimize candidate triggers. Backpropagation requires white-box access to the model, which is not available in our black-box setting. Therefore, gradient-based optimization techniques are not feasible. As a result, we employ a discrete search algorithm. We present the key components of our search algorithm.

\textbf{Search space}. Each type of attack is characterized by a number of parameters that constitute its search space. Given an attack type, we use these parameters to define its search space (Section~\ref{sec:attack-search-space}). We traverse this search space to find potential backdoor triggers.

\textbf{Search for target label}. We observe that a trigger that effectively targets one class is ineffective for other classes. As a result, we search for a trigger for each target label separately. 

\textbf{Simulated annealing (SA)}.
A range of search algorithms exist for discrete search problems. Some algorithms are ideal for convex optimization. However, stealthy backdoor attacks have non-convex (combinatorial) search spaces. SA is designed to operate in non-convex and discrete search problems~\cite{vanLaarhoven1987annealing}. It is an adaptation of the Metropolis–Hastings sampling algorithm~\cite{chib1995metro}, which is itself a Monte Carlo method. Several convergence guarantees are proven, showing that given enough iterations, SA can reach the global optima of the objective function~\cite{adams1994convergence, chib1995metro} in non-convex search spaces. This is not the case with some other discrete optimization techniques, such as Hill Climbing. Algorithm~\ref{alg:search} presents this algorithm in our context.

\textbf{Temperature schedule}. SA has a temperature schedule that balances exploration and exploitation. This schedule gradually reduces the temperature throughout the search process. An initially high temperature allows for exploration early in the process, while the eventually low temperature prioritizes exploitation late in the process~\cite{styblinski1990experiments}. To achieve this, $T$ is defined as:
\begin{equation}
\label{eq:temperature}
    T=\epsilon \cdot (\frac{1}{k+\epsilon}-\frac{1}{s+\epsilon})
\end{equation}
where $s$ is the total number of steps, $k$ is the current step, and $\epsilon$ is a parameter that controls how quickly the temperature drops. Early in the search, $k$ is low, and therefore $T$ is high, so the search will make suboptimal moves to explore the search space. As the search progresses and $k$ increases, $T$ drops and thus the search prioritizes making fewer mistakes and optimizing the best trigger.
The key advantage of SA over gradient-based optimizations is that SA ensures exploration in addition to exploitation.

\subsection{Attack Search Space}
\label{sec:attack-search-space}

Each attack type has a generic search space. Our detection technique is applicable to a wide range of attacks by expressing the search space of each attack type. This subsection gives four examples of how the search space of an attack can be expressed.

\textbf{Patch} attacks follow the backdoor definition presented in Neural Cleanse~\cite{wang2019cleanse}:
$$\mathcal{T}(x,\Delta=(p,m,\phi)) =x'$$
\begin{equation}
\label{eq:general}
x'_{i,j,c} =(1-m_{i,j})\cdot x_{i,j,c}+m_{i,j}\cdot p_{i,j,c}
\end{equation}
where $\mathcal{T}$ applies the trigger $\Delta$ to a clean sample $x \in \mathcal{X}$, producing the backdoored sample $x'$. The 3D tensor $p \in \left[0,1\right]^{c\times w\times h}$ contains the pixel intensity values representing the trigger pattern. The matrix $m \in \{0,1\}^{w\times h}$ contains values that determine which pixels of $x$ are overwritten by $p$. The result is a continuous patch of pixels stamped onto the original image (see Figure~\ref{fig:attack-types}-b). Thus, the search space is defined as the location, size, and pattern of the patch.

\textbf{Blended} attacks use the same backdoor function shown in Equation~\ref{eq:general}. The main difference is the matrix $m \in \left[-1,1\right]^{w\times  h}$, i.e. the trigger is blended into the background of the entire original image (see Figure~\ref{fig:attack-types}-c). Thus, the search space is the pattern and magnitude of blending for each pixel.

\textbf{Warped} attacks rely on backdoor function $\mathcal{T}$ that applies the warping field $M$ to a sample~\cite{nguyen2021wanet}. The warping field $M$ for an attack is randomly initialized as:
\begin{equation}
\label{eq:wanet}
    \begin{split}
    M=\omega(\uparrow(\psi(\mathrm{rand}_{[-1,1]}(k,k,2))\times s))
    \end{split}
\end{equation}
where $k$ is the control grid size, $s$ is the warping strength, $\mathrm{rand}_{[-1,1]}(.)$ is a random tensor generator with values in $[-1, 1]$, $\psi$ is a normalization function, $\uparrow$ is an upsampling step, and $\omega$ is a clipping function. The result is a warping field that is applied across the image and more difficult to detect visually (see Figure~\ref{fig:attack-types}-e). Thus, the search space is the initialization of the warping effect. 

\textbf{LIRA} attacks learn a transformation function $T_{\xi^{*}}$ which generates a noise trigger given a sample, by solving the following optimization~\cite{doan2021lira}:
\begin{equation}
\label{eq:lira}
    \begin{alignedat}{3}
      & \min_{\theta}& &\sum_{i=1}^{N} & &\alpha \mathcal{L}(f_{\theta}(x_{i}),l_{i})+\beta \mathcal{L}(f_{\theta}(T_{\xi^{*}(\theta)}(x_{i})), \phi(l_{i})) \\
      & s.t. \quad & &(i) & & \xi^{*}=\argmin_{\xi} \sum_{i=1}^{N}\mathcal{L}(f_{\theta}(T_{\xi}(x_{i})),\phi (l_{i})) \\
      & & &(ii) & & d(T(x),x) \leq \epsilon
    \end{alignedat}
\end{equation}    
where $f_{\theta}$ is the victim model, $T_{\xi^{*}}: \mathcal{X} \rightarrow \mathcal{X}$ is the transformation function that transforms a clean image into a backdoored image, $\alpha$ and $\beta$ control the mixed loss from clean and poisoned data, and $d$ is a function that measures visual difference between samples. After training, $T_{\xi^{*}}$ is used to generate imperceptible noise, serving as the trigger (see Figure~\ref{fig:attack-types}-f). In this case, the search space consists of a noise pattern. We find that a fixed noise pattern achieves high detection performance.

\section{Evaluation}
\label{sec:evaluation}

\subsection{Experimental Setup}

\textbf{Backdoor attack types}: We consider eight diverse attacks from the literature: patch~\cite{gu2017badnets}, blended~\cite{chen2017targeted}, filter~\cite{liu2019abs}, WaNet~\cite{nguyen2021wanet}, LIRA~\cite{doan2021lira}, MMS-BD~\cite{kwon2022multi}, c-BaN~\cite{salem2022dynamic}, and natural~\cite{fu2023eagle}. The descriptions of these attacks are discussed in Section~\ref{sec:related-work-backdoor-attacks}. These attacks employ a range of target label functions: \textbf{All2One} attacks misclassify samples from all classes to a single target class, \textbf{All2All} attacks misclassify samples from each source class to a corresponding target class, and \textbf{One2One} attacks misclassify samples from a single source class to a single target class. We evaluate DeBackdoor against the above variations. For the specific implementation details of these attacks, please refer to appendix~\ref{appendix:A}.

\textbf{Backdoor detection baselines}: We compare our technique with AEVA~\cite{guo2021aeva} and B3D~\cite{dong2021black}, the two backdoor detection techniques in the literature that operate in our pre-deployment, data-limited, single-instance, and black-box setting. To further benchmark the performance of our technique, we also compare against a range of techniques that operate in less restricted defense settings (post-deployment, data-available, multi-instance, white-box). These techniques include: ABS~\cite{liu2019abs}, Neural Cleanse (NC)~\cite{wang2019cleanse}, MNTD~\cite{xu2021mntd}, FreeEagle~\cite{fu2023eagle}, DF-TND~\cite{wang2020datafree}, STRIP~\cite{gao2019strip}, and ANP~\cite{wu2021adversarial}. For the Trojan Detection Challenge (TDC)~\cite{neurips2022tdc}, we also compare against all submissions made to the competition.

\textbf{Evaluation data}: For patch and blended attacks, we evaluate DeBackdoor against the TDC dataset containing 2000 clean/backdoored models which include CNN-5
trained on MNIST~\cite{deng2012mnist}, Wide ResNet-40-2~\cite{he2016deep} trained on CIFAR-10 /CIFAR-100~\cite{krizhevsky2009cifar, zagoruyko2016wide}, and Vision Transformers~\cite{beyer2022vit} trained on GTSRB~\cite{stallkamp2012gtsrb}. For WaNet, LIRA, MMS-BD, and c-BaN, we train 1700 clean/backdoored models using their code. These models include CNN-5 trained on MNIST and Pre-activation ResNet-18 trained on CIFAR-10 and GTSRB. For comparison with the out-of-setting baselines, we train 1700 clean/backdoored models using their code~\cite{fu2023eagle}. These models include CNN-7~\cite{lin2020composite} trained on MNIST, VGG-16~\cite{simonyan2015vgg} trained on CIFAR-10, GoogLeNet~\cite{szegedy2014goingdeeperconvolutions} trained on GTSRB, and ResNet-50 trained on ImageNet-R. ImageNet-R contains 20 classes from the ImageNet dataset~\cite{deng2009imagenet} (see Table~\ref{tab:imagenetr}).

\textbf{Tasks and evaluation metrics:} We consider three tasks: backdoor detection, target label prediction, and trigger synthesis. For detection, we evaluate area under ROC curve (AUROC) and true/false-positive rate (TPR/FPR). For target label prediction, we measure total accuracy (ACC) of predicted target labels. For trigger synthesis, we measure the intersection over union (IOU) and cosine similarity (CS) with respect to the ground truth.

\begin{figure}[t]
  \centering
  \vspace{-10mm}
  \begin{minipage}[htp]{0.35\textwidth}
    \centering
    \includegraphics[width=1\textwidth]{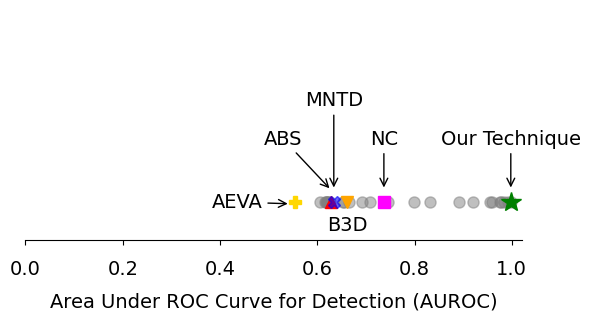}\\
      \end{minipage}
      \centering
  \begin{minipage}[htp]{0.35\textwidth}
    \centering
    \includegraphics[width=1\textwidth]{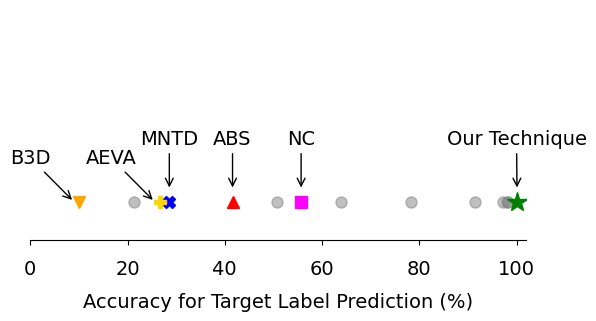}\\
      \end{minipage}
    
   \begin{minipage}[htp]{0.35\textwidth}
    \centering
    \includegraphics[width=1\textwidth]{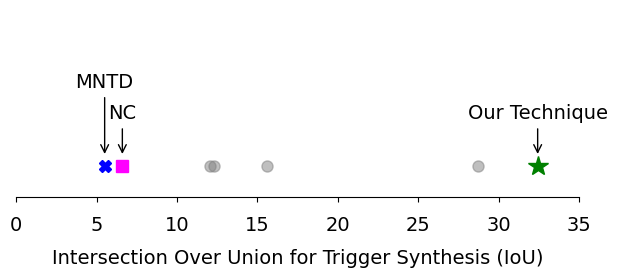}\\
      \end{minipage}
  \caption{A comparison of our detection technique to baselines and all submissions made to the Trojan Detection Challenge (TDC)~\cite{neurips2022tdc} across the three evaluation tasks.   }
  \label{fig:tdc}
\end{figure}

\begin{figure*}[t]
  \centering    \includegraphics[width=0.75\textwidth]{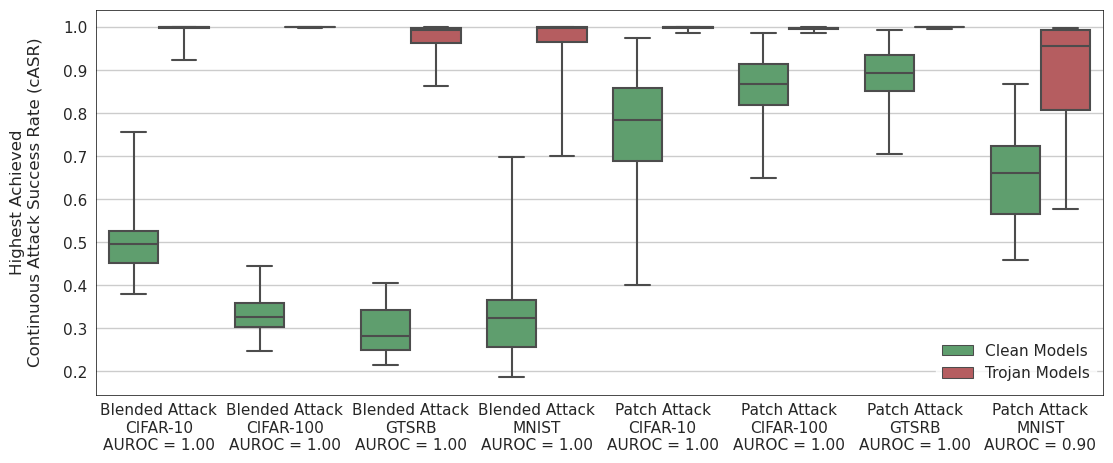}
  \caption{The continuous Attack success rate (cASR) detection scores across different attack techniques and datasets. Given each dataset and each attack technique, 125 clean and 62 backdoored models are used for measuring detection performance.}   \label{fig:score-distributions}
\end{figure*}

\textbf{Benchmarks}. Throughout the literature, dozens of attack and defense techniques are proposed~\cite{li2022survey}. Comparing different techniques is difficult because: (1) Each paper has evaluated with different scenarios, attack types, baselines, and data, and (2) many papers do not publish/maintain code implementations. Public competitions, such as the Trojan Detection Challenge (TDC) that was hosted at NeurIPS~\cite{neurips2022tdc}, aim to tackle this problem by providing a standard scenario and evaluation data. Therefore, Section~\ref{sec:TDC} focuses on experiments on TDC, which covers patch and blended attacks. We evaluate against dynamic attacks (WaNet, LIRA, c-BaN) in Section~\ref{sec:dynamic}. There are very few defense techniques in the literature with defense settings as restrictive as ours. Therefore, we extend our comparison to out-of-setting techniques in Section~\ref{sec:outofsetting}. However, it is notable that out-of-setting techniques are not applicable to the defense scenario discussed in this paper. Finally, we ablate the parameters of DeBackdoor in Section~\ref{sec:ablation}.

\begin{figure}[t]
  \centering    \includegraphics[width=0.45\textwidth]{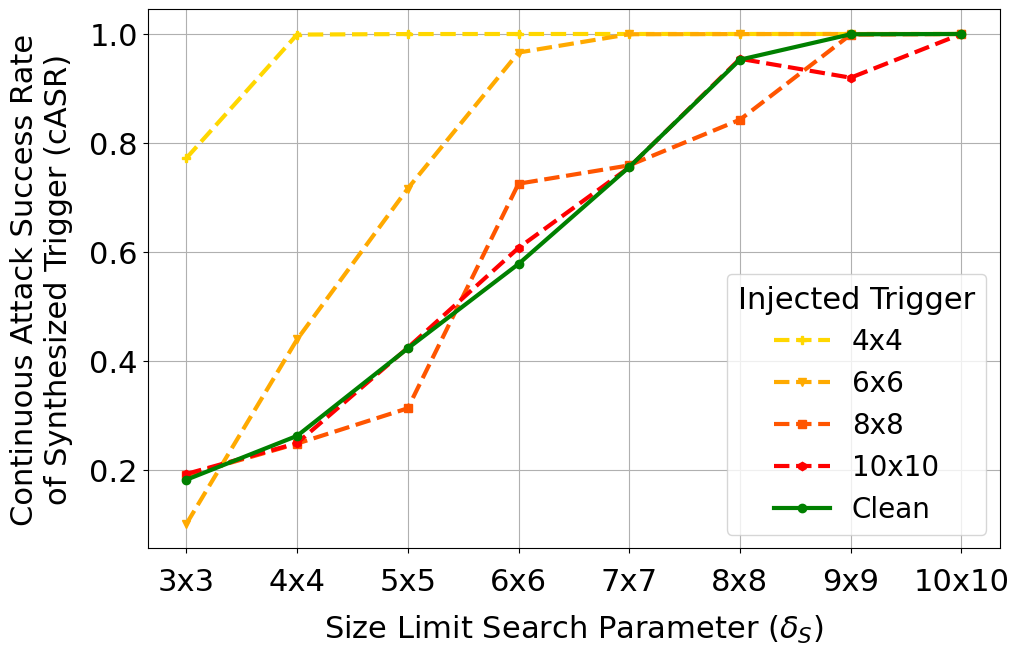}
  \caption{A comparison of the Continuous Attack Success Rate (cASR) of synthesized triggers for models injected with triggers of various sizes and DeBackdoor run with different settings of the size limit search parameter $\delta_{S}$.}
  \label{fig:stronger-triggers}
\end{figure}

\subsection{Trojan Detection Challenge}
\label{sec:TDC}

We evaluate DeBackdoor on the TDC dataset~\cite{neurips2022tdc}. Figure~\ref{fig:tdc} shows that our technique outperforms AEVA and B3D, as well as the three white-box/multi-instance baselines (ABS, Neural Cleanse, MNTD), and all of the 39 submissions made to the challenge, in all three tasks. The highest-scoring submission operates by recording the parameters of 1000 clean/backdoored models in the competition train set and training a classifier (white-box, multi-instance). The techniques and code for all other submissions are not published.

\begin{figure*}[t]
  \centering
  \begin{minipage}[htp]{0.3\textwidth}
    \centering
    \hspace{-8mm}\includegraphics[width=1\textwidth]{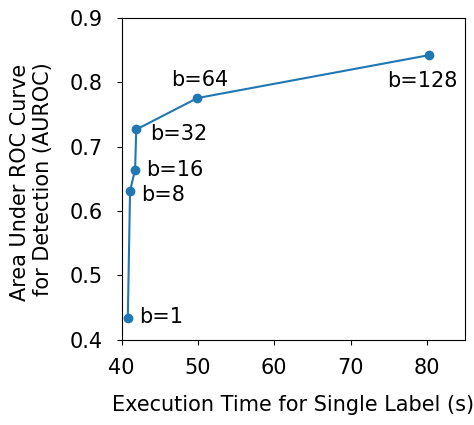}  \\
    (a) Effect of search batch size on Area Under ROC Curve (AUROC)
  \end{minipage}
  \hspace{3mm}
    \centering
    \hspace{1mm} \begin{minipage}[htp]{0.3\textwidth}
    \centering
    \hspace{-8mm}\includegraphics[width=1\textwidth]{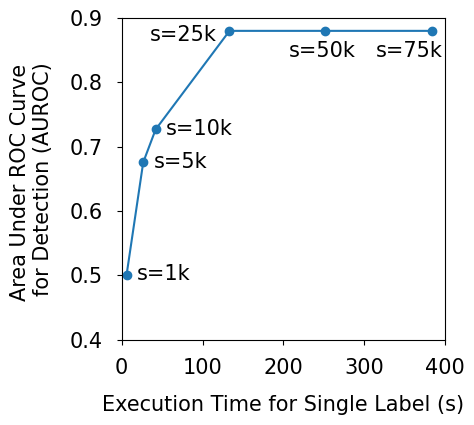}\\
    (b) Effect of number of Simulated Annealing steps on AUROC
  \end{minipage}
  \hspace{3mm}
 \begin{minipage}[htp]{0.3\textwidth}
    \centering    \hspace{-5mm}\includegraphics[width=1\textwidth]{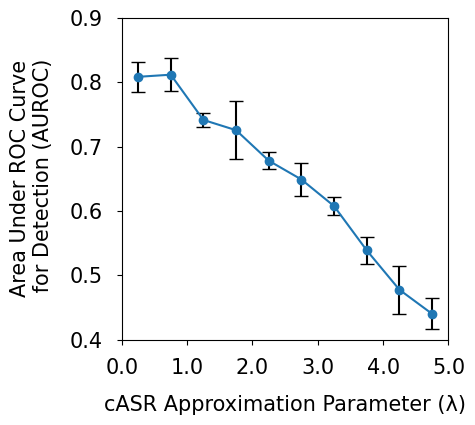}\\
    (c) Effect of the choice of parameter $\lambda$ on AUROC
  \end{minipage}
  \caption{A study of the effect of batch size \textbf{b}, number of Simulated Annealing (SA) steps \textbf{s}, and the choice of $\lambda$ on Area Under ROC Curve (AUROC) to separate clean/backdoor models. To search for $b$ and $\lambda$, we used sub-optimal $s$, and to search for $s$ we used suboptimal $b$ to avoid AUROC $=1.0$ and ensure that the comparison is meaningful. We found that $b=32$, $s=10000$, and $\lambda=0.6$ achieve an optimum speed-accuracy trade-off for our experiments. For LIRA and WaNet, $s=1000$ is sufficient.}
  \label{fig:hyper-parameters}
\end{figure*}

The reason that DeBackdoor outperforms AEVA and B3D can be partially attributed to the use of the cASR proxy. Figure~\ref{fig:hyper-parameters}-c shows that as $\lambda$ increases and cASR becomes less continuous (i.e. similar to the optimization performed in AEVA and B3D), AUROC drops significantly. Another reason is the use of SA in DeBackdoor, which can avoid local minima better~(Figure~\ref{fig:hyper-parameters}-b).

It is notable that white-box access to the model is fundamentally helpful because it is useful to study the behavior of the model (e.g. by calculating gradients). Having access to a set of clean/backdoored models is also helpful as the defender can train a classifier. The reason that DeBackdoor outperforms current white-box and multi-instance techniques is not because white-box access or access to a set of reference models hurt, but rather because of the following reasons:
\begin{itemize}
\item \textbf{Realistic assumptions}: These white-box and multi-instance techniques make assumptions about the backdoor mechanism and trigger that do not hold in all attack scenarios, thus failing to detect all of the backdoored models. For example, ABS~\cite{liu2019abs} assumes that the trigger is learned by a small subset of compromised neurons, which fails in TDC. Neural Cleanse~\cite{wang2019cleanse} assumes that backdoored models are detected when the candidate trigger synthesized for one class is significantly smaller than the triggers synthesized for all other classes. This assumption also doesn't hold in TDC. MNTD~\cite{xu2021mntd} creates a synthetic training dataset of clean/backdoored models and uses this dataset to train a meta-classifier. Thus, MNTD relies on the assumption that its distribution of synthetic models is similar to the distribution of actual attacked models, which is again shown to not be the case. 

\item \textbf{White-box vs black-box access to the model}: Even though black-box techniques do not have direct access to model parameters, as discussed in Section~\ref{sec:search-algo}, there are mathematical guarantees that Simulated Annealing (which is compatible with black-box access) can sufficiently explore the search space and reach global optima.

\item \textbf{Deductive solution}: Our technique focuses on the basic assumption that backdoored models must contain a trigger with high ASR. This assumption cannot be broken easily because it is a core principle of all backdoor attacks. We rely on this principle and perform an extensive search to generate effective triggers. Since our model generates triggers, it has a low chance of false positives.
\end{itemize}

\textbf{Detection}. For each model architecture and image dataset, our technique achieves perfect AUROC for detecting blended attacks (Figure~\ref{fig:score-distributions}). For patch attacks, we achieve perfect AUROC for detecting Wide ResNets trained on CIFAR-10/CIFAR-100 and ViTs trained on GTSRB. However, our AUROC drops from $1.0$ to $0.9$ for CNNs trained on MNIST.

\begin{figure}[t]
  \centering
  \begin{minipage}[htp]{0.2\textwidth}
    \centering
    \includegraphics[width=1\textwidth]{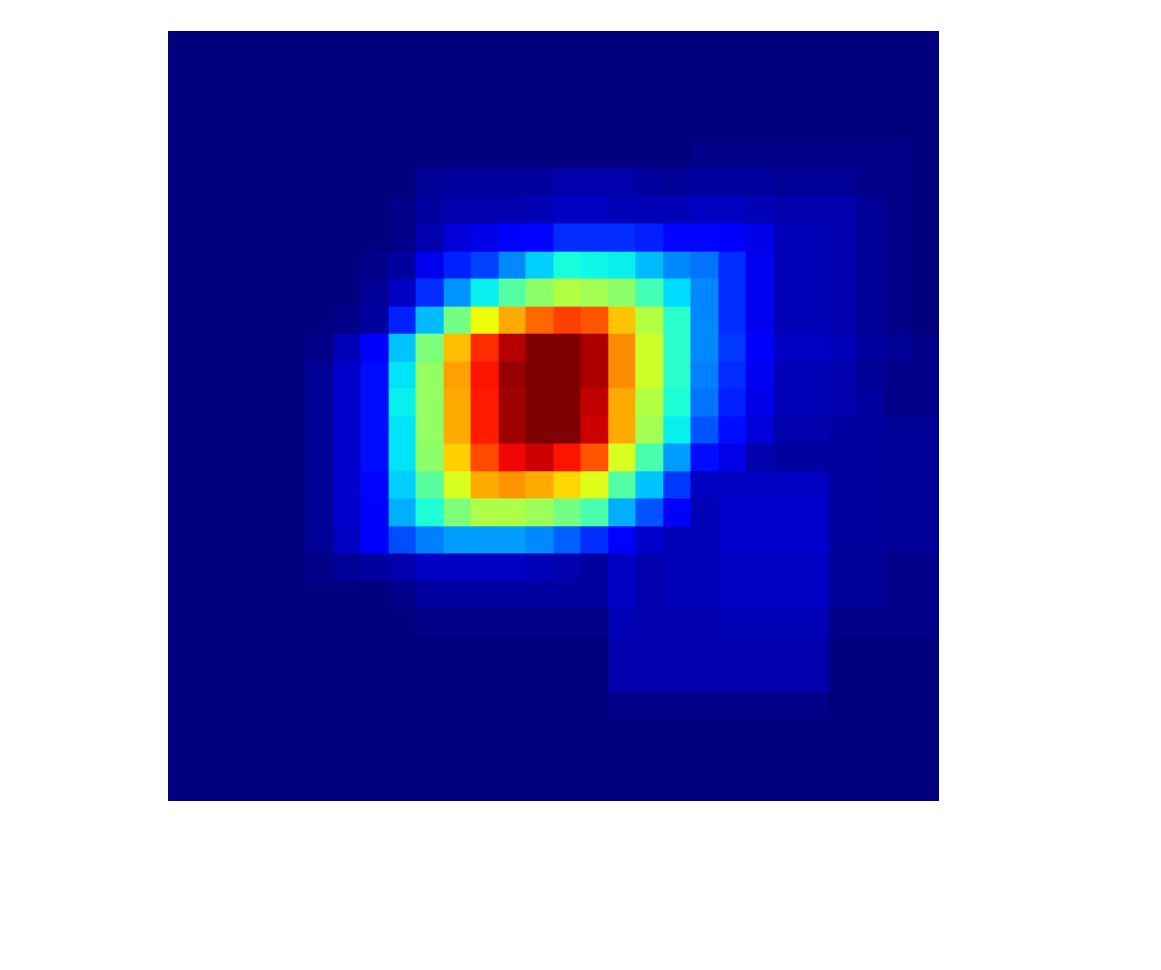}
  \end{minipage}
  \hspace{-10mm}
    \centering
  \begin{minipage}[htp]{0.2\textwidth}
    \centering
    \includegraphics[width=1\textwidth]{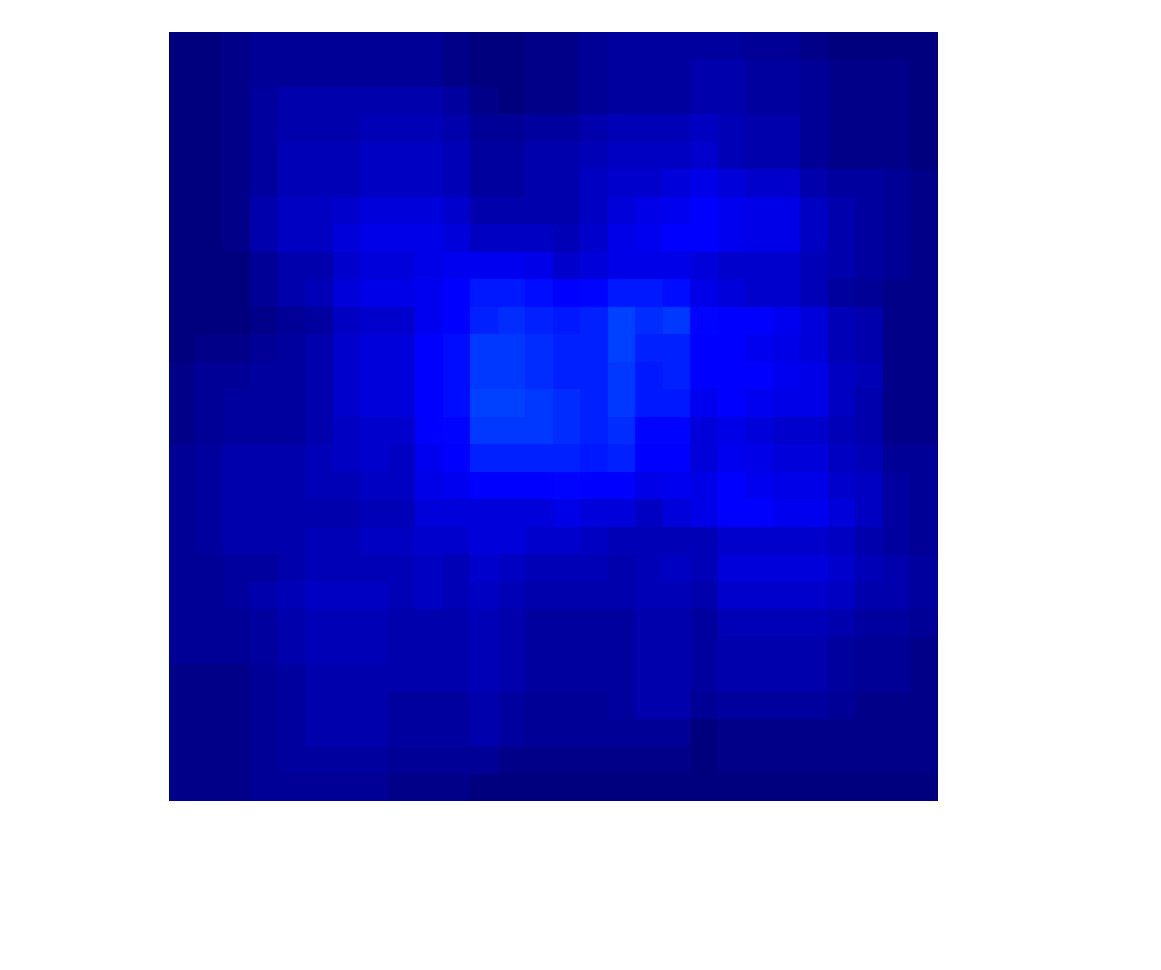}
  \end{minipage}
            \caption{\textbf{Left}: A heatmap of patch triggers generated for clean CNNs trained on MNIST. \textbf{Right}: A heatmap of patch triggers generated for backdoored CNNs trained on MNIST.}   \label{fig:tough}
\end{figure}

The drop in detection performance on the MNIST dataset is because the TDC contains patch triggers as large as $10 \times 10$ pixels. These patches are large enough to rewrite the digit in the simple $28 \times 28$ MNIST images, leading to a drop in AUROC. This is further verified by the fact that triggers synthesized for clean models are densely located in the center of the image where the digit is located (Figure~\ref{fig:tough}). Triggers that are too strong are unreasonable because they resemble rewriting the input. Therefore, Attack Success Rate (ASR) for even clean models approaches $100\%$, as shown in Figure~\ref{fig:stronger-triggers}. 

\begin{figure}[t]
  \centering    \includegraphics[width=0.22\textwidth]{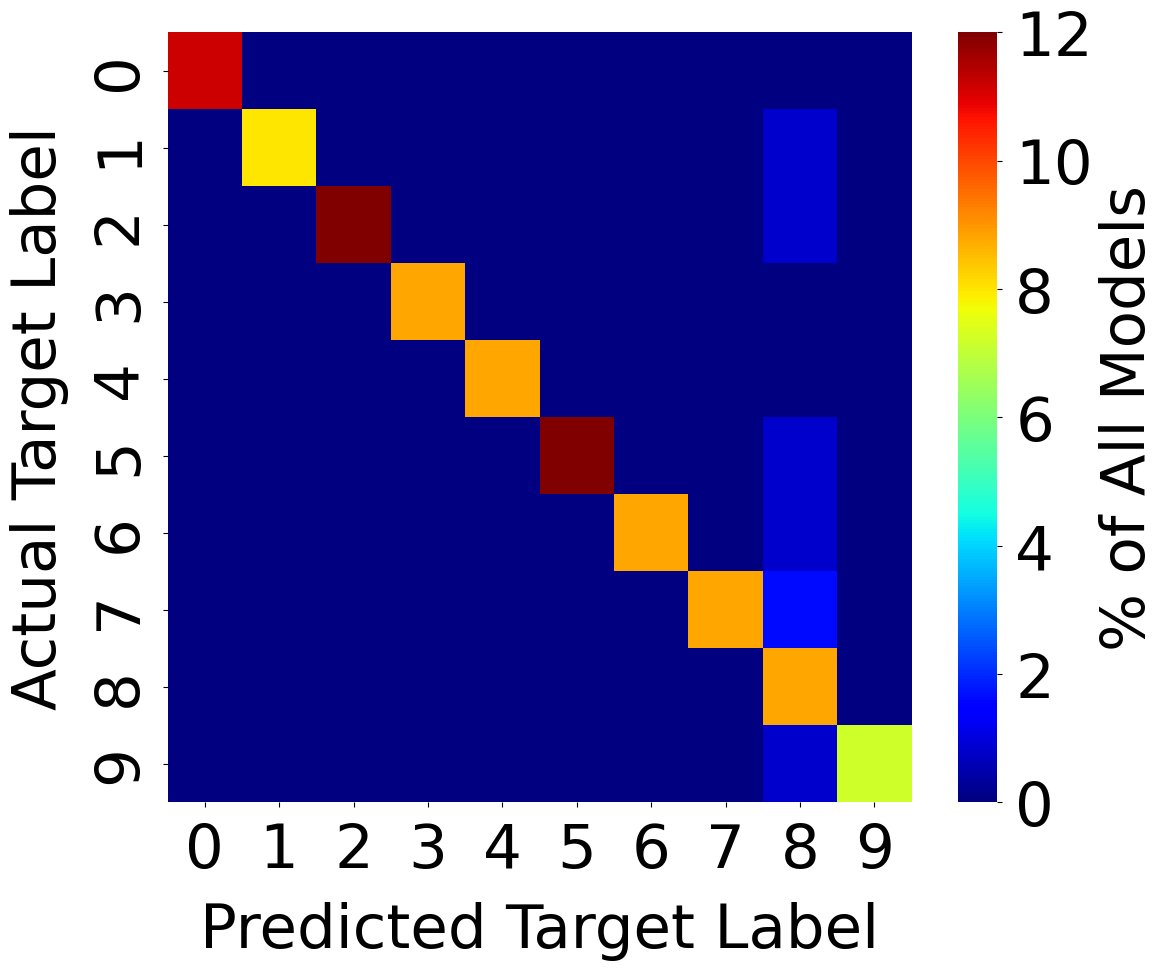}
  \caption{The confusion matrix of target label predictions made for patch-attacked models trained on MNIST.}   \label{fig:confusion-matrix}
\end{figure}

\textbf{Target label prediction}. For each model architecture and image dataset in TDC, our technique achieves perfect accuracy for predicting the target label of all blended attacks. For patch attacks, we achieve perfect accuracy for predicting the target label of Wide ResNets trained on CIFAR-10/CIFAR-100 and ViTs trained on GTSRB. However, our accuracy once again drops from $100\%$ to $94.4\%$ for patch-attacked CNNs trained on MNIST. Figure~\ref{fig:confusion-matrix} reveals that all of the misclassified target labels belong to class 8. Here, false positives arise due to natural backdoors present in the data classes; class 8 is vulnerable because samples from other classes can be easily altered to share the natural features of the digit 8. This phenomenon is known as a ``natural'' backdoor \cite{tao2022natural}.

\begin{figure*}[t]
    \centering
    \begin{minipage}[htp]{0.22\textwidth}
        \centering
        \hspace{-8mm}\includegraphics[width=1\textwidth]{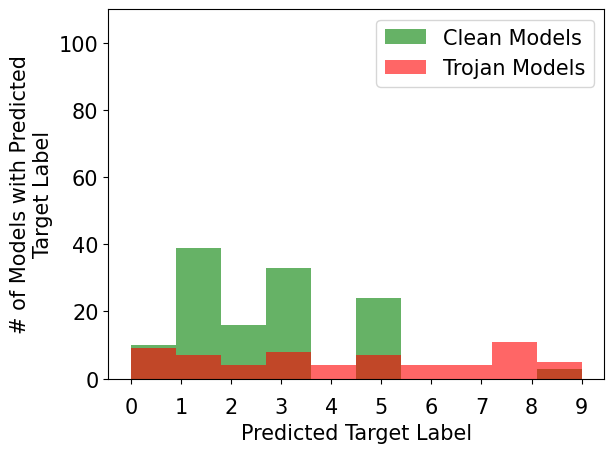}  \\
        (a) CIFAR-10
    \end{minipage}
    \begin{minipage}[htp]{0.22\textwidth}
        \centering
        \hspace{-8mm}\includegraphics[width=1\textwidth]{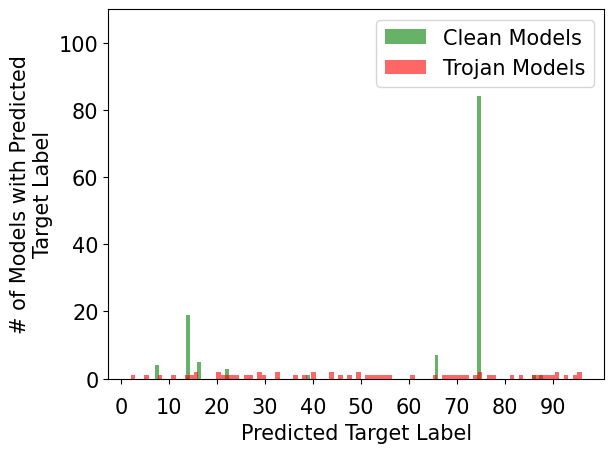}\\
        (b) CIFAR-100
    \end{minipage}
    \begin{minipage}[htp]{0.22\textwidth}
        \centering
        \hspace{-8mm}\includegraphics[width=1\textwidth]{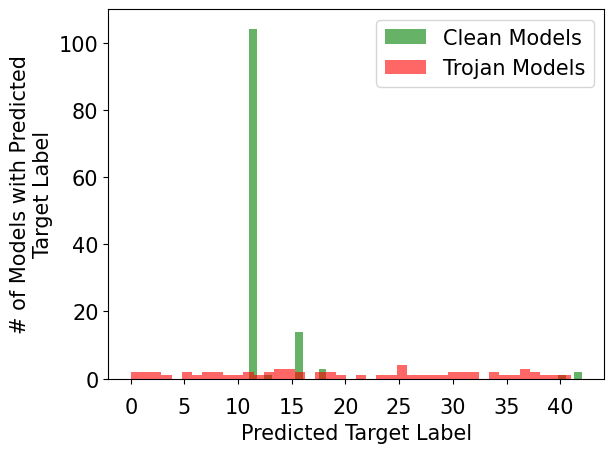}  \\
        (c) GTSRB
    \end{minipage}  
    \begin{minipage}[htp]{0.22\textwidth}
        \centering
        \hspace{-8mm}\includegraphics[width=1\textwidth]{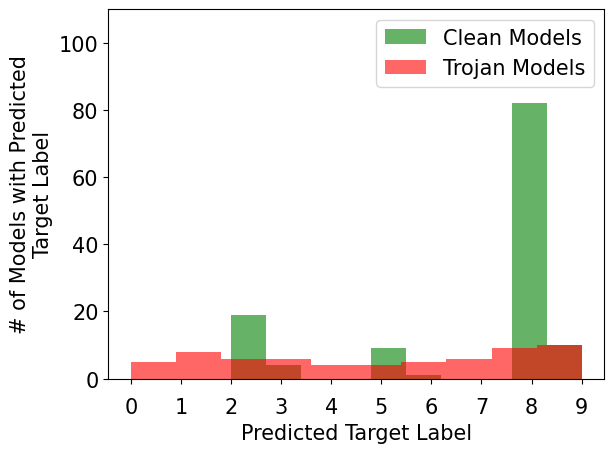}\\
        (d) MNIST
    \end{minipage}
    \caption{A comparison of the distributions of predicted target labels for clean/backdoored models trained on four TDC datasets.}
    \label{fig:label-preds}
\end{figure*}

In fact, we observe the presence of vulnerable classes across the four datasets by analyzing the target label predictions made for clean and backdoored models (Figure~\ref{fig:label-preds}). For backdoored models, the predictions are uniformly distributed as the attacker randomly selects the target label and our technique correctly predicts this target. However, triggers synthesized for clean models have a bias towards one class (class 75 in Figure~\ref{fig:label-preds}-b, class 11 in Figure~\ref{fig:label-preds}-c, and class 8 in Figure~\ref{fig:label-preds}-d). These classes are naturally vulnerable to patch attacks since samples from other classes can be easily modified to appear similar to samples in the vulnerable class.

\begin{figure*}[]
  \centering    \includegraphics[width=0.8\textwidth]{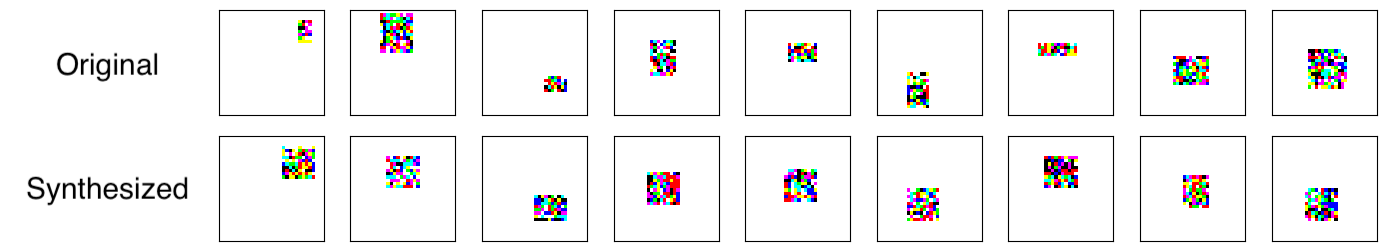}
  \caption{A visualization of the original triggers and the corresponding triggers synthesized by our technique for vision transformers trained on the GTSRB dataset.}
  \label{fig:trigger-visualizations}
  \vspace{-4mm}
\end{figure*}

\textbf{Trigger synthesis}. Our technique outperforms the baselines and competition submissions in the task of trigger synthesis. Figure~\ref{fig:trigger-visualizations} presents a few examples of patch triggers injected into the model and our synthesized triggers.

\subsection{Dynamic and Complex Attacks}
\label{sec:dynamic}

Dynamic attacks are designed to be especially stealthy by injecting backdoors that cause the model to act maliciously in the presence of sample-specific triggers. Additionally, some of the attacks that we consider in this section have All2All variations, whereby samples from different source classes are mislabeled to corresponding attacker-chosen target classes. These attack characteristics aim to break detection techniques that rely on the universality of the trigger, i.e. finding a single trigger that causes all samples to be misclassified to a single target class. In practice, although these attacks generate sample-specific triggers, the process for generating these triggers is fixed. For example, in the WaNet attack, a fixed set of parameters is used to generate the warping field that is then applied as a trigger to clean inputs. Our technique leverages this by searching through the fixed parameters of the trigger generation function and identifying the configuration that yields the attacker's sample-specific triggers.

In addition, while the primary goal of dynamic attacks is to train the victim model to misclassify samples that contain sample-specific triggers, we observe that the victim model also learns triggers that are sample-agnostic as a byproduct of the backdooring process. This is verified by extensive results that show our technique is able to find triggers which achieve high ASR when applied to any sample, for models that were injected with sample-specific triggers (e.g. LIRA). DeBackdoor detects dynamic attacks by finding these byproduct triggers.

In the following, we discuss the details of the different configurations that our technique employs to achieve high detection performance (Table~\ref{tab:auroc-dynamic}) against the All2One and All2All variations of the WaNet and LIRA dynamic attacks.

\begin{table}[h]
\caption{The Area Under Receiver Operator Curve (AUROC) for detection of WaNet~\cite{nguyen2021wanet} and LIRA~\cite{chen2017targeted} attacks (across two label strategies and three different datasets) by AEVA~\cite{guo2021aeva}, B3D~\cite{dong2021black}, and our technique.}
\small
\centering
\begin{tabular}{ll l l l l}
\hline
\multirow{2}{*}{Attack} & Label & \multirow{2}{*}{Dataset} & \multicolumn{3}{c}{Detection Technique} \\ 
\cline{4-6}
 & Strategy & & AEVA & B3D & Ours \\
\hline
\multirow{6}{*}{WaNet}
  & \multirow{3}{*}{All2One} & CIFAR-10 & 0.58 & 0.59 & 1.00 \\
  &                          & GTSRB    & 0.56 & 0.57 & 1.00 \\
  &                          & MNIST    & 0.53 & 0.81 & 1.00 \\
\cline{2-6}
  & \multirow{3}{*}{All2All} & CIFAR-10 & 0.53 & 0.53 & 1.00 \\
  &                          & GTSRB    & 0.47 & 0.65 & 1.00 \\
  &                          & MNIST    & 0.53 & 0.79 & 1.00 \\
\hline
\multirow{6}{*}{LIRA}
  & \multirow{3}{*}{All2One} & CIFAR-10 & 0.79 & 0.53 & 1.00 \\
  &                          & GTSRB    & 0.78 & 0.81 & 1.00 \\
  &                          & MNIST    & 0.68 & 0.78 & 1.00 \\
\cline{2-6}
  & \multirow{3}{*}{All2All} & CIFAR-10 & 0.55 & 0.53 & 1.00 \\
  &                          & GTSRB    & 0.55 & 0.78 & 1.00 \\
  &                          & MNIST    & 0.56 & 0.84 & 1.00 \\
\hline
\end{tabular}

\label{tab:auroc-dynamic}
\end{table}

\textbf{WaNet}. To detect the WaNet~\cite{nguyen2021wanet} attack, we use a search space according to Equation~\ref{eq:wanet}. We search over the control grid size $k \in \{3,4,5,6\}$, according to the original paper. For each $k$, we search for the $2 \times k \times k$ secret control grid. Table~\ref{tab:auroc-dynamic} shows that we achieve perfect detection accuracy. The first row in Table~\ref{tab:dynamic-partial-casr} shows that our technique separates clean/WaNet-attacked models by a wide margin. We use the standard threshold of cASR $>95\%$, but note that any threshold between $49\%$ and $99\%$ yields perfect accuracy for all datasets. Meanwhile, AEVA performs near-random, and B3D only achieves nonrandom performance on the simplest dataset (MNIST). This is because both of these techniques leverage generic optimization schemes that do not apply in the case where a warping-based technique is used to create distinct triggers per sample.

\textbf{LIRA}. To detect the LIRA~\cite{doan2021lira} attack, we observe that the noise generated by $T_{\xi^{*}}$ (Equation~\ref{eq:lira}) has a smaller norm than the perturbations applied in other attacks because $T_{\xi^{*}}$ is optimized to minimize it. Therefore, we used the search space for blended attacks with a tighter perturbation bound on the noise to detect LIRA triggers. Table~\ref{tab:auroc-dynamic} shows perfect detection accuracy. The second row of Table~\ref{tab:dynamic-partial-casr} shows that the margin between clean and backdoor models is also wide ($>59\%$). AEVA and B3D achieve higher detection scores on LIRA than WaNet because of the greater presence of attack byproducts, in the form of static triggers, that can be captured by their optimization. However, their Hill Climbing approaches fall into local minima, while Simulated Annealing allows our technique to reach global optima and thus perform better.

\begin{table}[t]
\small
\setlength{\tabcolsep}{4pt}
\caption{The mean cASR of synthesized triggers for clean models, models attacked by WaNet with All2One/All2All strategies, and models attacked by LIRA with All2One/All2All strategies.}
\label{tab:dynamic-partial-casr}
\centering
\begin{tabular}{ll l c c c}
\toprule
\multirow{2}{*}{Attack} & Label & \multirow{2}{*}{Category} 
                        & \multicolumn{3}{c}{Dataset} \\
\cline{4-6}
 & Strategy & & CIFAR-10 & GTSRB & MNIST \\
\midrule
\multirow{4}{*}{WaNet} 
 & \multirow{2}{*}{All2One} 
    & Clean    & 28.8 & 46.9 & 48.1 \\
 & 
    & Backdoor & 99.9 & 99.8 & 99.9 \\
\cline{2-6}
 & \multirow{2}{*}{All2All} 
    & Clean    & 28.8 & 46.9 & 48.1 \\
 & 
    & Backdoor & 97.8 & 99.7 & 99.5 \\
\midrule
\multirow{4}{*}{LIRA} 
 & \multirow{2}{*}{All2One} 
    & Clean    & 15.8 & 10.2 & 16.0 \\
 & 
    & Backdoor & 75.0 & 93.4 & 98.6 \\
\cline{2-6}
 & \multirow{2}{*}{All2All} 
    & Clean    & 15.8 & 10.2 & 16.0 \\
 & 
    & Backdoor & 67.8 & 59.3 & 97.5 \\
\bottomrule
\end{tabular}
\end{table}

To tackle the All2All variations of the WaNet and LIRA attacks, we make a simple modification to our technique; we run the search algorithm using a set of validation images that all belong to a single source class $s$. If a trigger that is specific to this source class exists, it misclassifies all samples from class $s$ to their target class $t=\phi(s)$. Hence, the trigger with the highest cASR among all pairs of $(s,t)$ is the most likely candidate. Otherwise, the model is clean. Table~\ref{tab:auroc-dynamic} shows that our technique achieves perfect detection scores across all datasets and outperforms the baselines. We highlight that in practice, AEVA and B3D operate by synthesizing triggers for each class and using outlier detection to identify whether any one of them is significantly smaller than the rest, only flagging the model as backdoored if such a trigger is found. In All2All attacks, many classes are infected and thus small triggers can be synthesized for more than one label, failing the aforementioned outlier assumption. In contrast, our technique does not depend on outlier detection and examines each result independently via the cASR, making it adaptable to a broad range of attack types.

\begin{figure}[t]
  \centering
  \begin{minipage}[htp]{0.47\textwidth}
    \centering
    \includegraphics[width=1\textwidth]{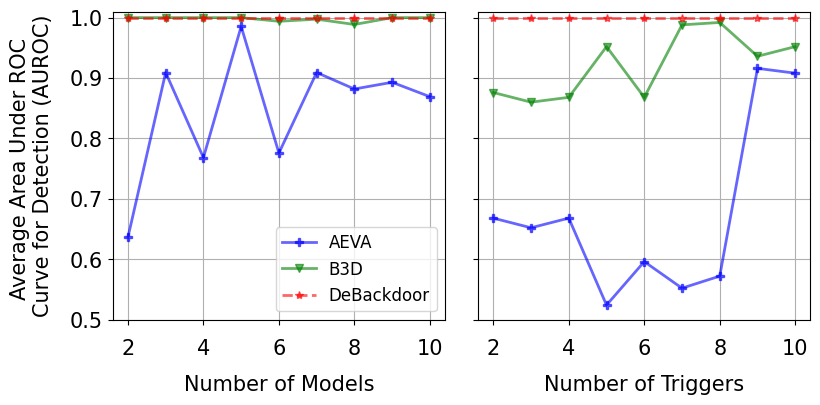}
  \end{minipage}
  \caption{\textbf{Left}: A comparison of detection performance against the MMS-BD~\cite{kwon2022multi} attack in settings with varying numbers of models, where each model is injected with a distinct backdoor. \textbf{Right}: A comparison of detection performance against the c-BaN~\cite{salem2022dynamic} attack with varying numbers of triggers, where each model is injected with $n$ different triggers.}   \label{fig:multi-auroc}
\end{figure}

Next, we explore the effectiveness of DeBackdoor against two recent attacks that target multiple models~\cite{kwon2022multi} and inject multiple triggers~\cite{salem2022dynamic}.

\textbf{Multi-Model Selective Backdoor Attack} (MMS-BD)~\cite{kwon2022multi} operates in settings with multiple models, inducing specific and distinct misclassification in each desired model by incorporating the locations of specific triggers in the attack. This is achieved by training each model to additionally learn backdoor samples that are incorrectly classified according to a specific trigger position for each model. The baselines and our method are effective in detecting this attack (Figure~\ref{fig:multi-auroc}). This is because the attack uses small patch triggers and All2One label strategy, making it possible to detect using detection techniques which are agnostic to the location and selective nature of the backdoor.

\textbf{Conditional Backdoor Generating Network (c-BaN)}~\cite{salem2022dynamic} uses a generative network to automatically construct multiple sample-specific triggers with different target labels. The generative network is trained jointly with the backdoor model, taking latent code sampled from a uniform distribution to generate a trigger and selecting a random location on the input, thereby making the trigger dynamic in terms of pattern and locations. Figure~\ref{fig:multi-auroc} shows that DeBackdoor successfully detects this attack across varying numbers of triggers injected into each model, while the baseline detection methods tend to achieve higher detection performance as the number of triggers increases. The main reason for this is that injecting many variations of triggers across multiple target classes into a single model makes detection easier as the attack leaves a large footprint on the model. Consequently, detection techniques only need to identify one of many injected triggers and target classes in order to successfully detect that the model is backdoored.

\begin{table}[t]
\small
\setlength{\tabcolsep}{4pt}
\caption{A comparison of different detection techniques against semantic backdoors that employ natural triggers. DeBackdoor outperforms all baselines by a large margin.}
\label{tab:natural-trigger}
\centering
\begin{tabular}{l c c c c}
\toprule
\multirow{2}{*}{Dataset} & Model & \multirow{2}{*}{Attack} & Detection & \multirow{2}{*}{TPR/FPR} \\
 & Architecture & & Technique & \\
\midrule
\multirow{6}{*}{ImageNet-R} & \multirow{6}{*}{ResNet-50} & \multirow{6}{*}{Natural} & ABS & 0.31/0.01 \\
& & & ANP & 0.10/0.05 \\
& & & DF-TND & 0.00/0.04 \\
& & & FreeEagle & 0.62/0.05 \\
& & & NC & 0.00/0.03 \\
& & & STRIP & 0.08/0.05 \\
& & & DeBackdoor & \textbf{1.00/0.00} \\
 
\bottomrule
\end{tabular}
\end{table}

\begin{table*}[t]
\small
\setlength{\tabcolsep}{4pt}
\caption{A comparison of different detection techniques on a number of datasets, architectures, and backdoor settings. Given each backdoor setting and defense technique, we report True Positive Rate (TPR) and False Positive Rate (FPR). Notice that DeBackdoor outperforms FreeEagle and other techniques in most tasks. Most performance numbers are borrowed from~\cite{fu2023eagle}.}
\label{tab:white-box}
\centering
\begin{tabular}{l c c c c c c c c}
\toprule
 & \multirow{4}{*}{Dataset} & & \multicolumn{6}{c}{Backdoor Settings \& TPR/FPR} \\
\cline{4-9}
Detection & & Model & \multicolumn{3}{c}{All2One} & \multicolumn{3}{c}{One2One} \\
\cline{4-9}
Technique & & Architecture & Patch & Blending & Filter & Patch & Blending & Filter\\
 & & & Trigger & Trigger & Trigger & Trigger & Trigger & Trigger\\
 \midrule
 \multirow{4}{*}{ABS} 
 & CIFAR-10 & VGG-16 & 0.37/0.04 & 0.61/0.05 & 0.21/0.04 & 0.56/0.05 & 0.25/0.02 & 0.26/0.05\\
 & GTSRB & GoogLeNet & 0.56/0.05 & 0.62/0.04 & 0.34/0.05 & 0.43/0.05 & 0.26/0.04 & 0.13/0.05\\
 & ImageNet-R & ResNet-50 & 0.67/0.05 & 0.22/0.01 & 0.73/0.03 & 0.43/0.05 & 0.40/0.04 & 0.32/0.05\\
 & MNIST & CNN-7 & 0.71/0.05 & 0.64/0.05 & 0.23/0.04 & 0.35/0.02 & 0.15/0.05 & 0.23/0.05\\
\midrule
\multirow{4}{*}{ANP} 
 & CIFAR-10 & VGG-16 & 0.90/0.01 & 0.76/0.04 & 0.77/0.03 & 0.62/0.05 & 0.51/0.05 & 0.57/0.05\\
 & GTSRB & GoogLeNet & 0.90/0.05 & 0.74/0.05 & 0.53/0.05 & 0.28/0.05 & 0.13/0.05 & 0.14/0.05\\
 & ImageNet-R & ResNet-50 & 0.99/0.05 & 0.96/0.03 & 0.74/0.05 & 0.31/0.05 & 0.23/0.05 & 0.19/0.05\\
 & MNIST & CNN-7 & 0.83/0.05 & 0.86/0.05 & 0.73/0.05 & 0.71/0.05 & 0.68/0.05 & 0.43/0.05\\
\midrule
\multirow{4}{*}{DF-TND} 
 & CIFAR-10 & VGG-16 & 0.00/0.02 & 0.00/0.04 & 0.00/0.03 & 0.00/0.04 & 0.01/0.03 & 0.03/0.05\\
 & GTSRB & GoogLeNet & 0.23/0.05 & 0.08/0.04 & 0.31/0.05 & 0.19/0.05 & 0.17/0.05 & 0.28/0.04\\
 & ImageNet-R & ResNet-50 & 0.76/0.05 & 0.32/0.05 & 0.90/0.03 & 0.18/0.05 & 0.23/0.05 & 0.38/0.05\\
 & MNIST & CNN-7 & 0.05/0.04 & 0.23/0.05 & 0.00/0.02 & 0.04/0.01 & 0.09/0.05 & 0.03/0.05\\
\midrule
\multirow{4}{*}{FreeEagle} 
 & CIFAR-10 & VGG-16 & 0.98/0.03 & 0.73/0.04 & 0.85/0.04 & 0.71/0.05 & 0.72/0.05 & 0.74/0.04\\
 & GTSRB & GoogLeNet & 0.99/0.03 & 0.99/0.04 & \textbf{1.00/0.03} & 0.89/0.03 & \textbf{0.76/0.04} & \textbf{0.84/0.05}\\
 & ImageNet-R & ResNet-50 & 0.99/0.04 & 0.86/0.03 & 0.99/0.02 & 0.74/0.03 & 0.73/0.04 & 0.78/0.05\\
 & MNIST & CNN-7 & 0.97/0.03 & 0.81/0.05 & 0.79/0.01 & 0.78/0.03 & 0.70/0.04 & 0.72/0.03\\
\midrule
\multirow{4}{*}{NC} 
 & CIFAR-10 & VGG-16 & 0.90/0.00 & 0.70/0.00 & 0.13/0.05 & 0.07/0.05 & 0.02/0.04 & 0.02/0.05\\
 & GTSRB & GoogLeNet & \textbf{1.00/0.00} & \textbf{1.00/0.00} & 0.51/0.05 & 0.21/0.05 & 0.33/0.05 & 0.04/0.05\\
 & ImageNet-R & ResNet-50 & 0.75/0.00 & 0.68/0.02 & 0.23/0.05 & 0.00/0.00 & 0.00/0.00 & 0.00/0.00\\
 & MNIST & CNN-7 & 0.83/0.00 & 0.90/0.00 & 0.32/0.02 & 0.23/0.05 & 0.13/0.05 & 0.28/0.02\\
\midrule
\multirow{4}{*}{STRIP} 
 & CIFAR-10 & VGG-16 & 0.89/0.04 & \textbf{0.92/0.04} & 0.10/0.03 & 0.00/0.02 & 0.04/0.05 & 0.02/0.05\\
 & GTSRB & GoogLeNet & 0.97/0.01 & 0.57/0.05 & 0.34/0.05 & 0.10/0.05 & 0.01/0.05 & 0.11/0.05\\
 & ImageNet-R & ResNet-50 & 0.44/0.05 & 0.53/0.05 & 0.14/0.05 & 0.10/0.05 & 0.03/0.02 & 0.07/0.03\\
 & MNIST & CNN-7 & 0.83/0.05 & 0.00/0.01 & 0.00/0.02 & 0.00/0.04 & 0.00/0.03 & 0.00/0.01\\
\midrule
\multirow{4}{*}{DeBackdoor} 
 & CIFAR-10 & VGG-16 &\textbf{1.00/0.00} & 0.90/0.04 & \textbf{1.00/0.00} & \textbf{0.88/0.04} & \textbf{0.97/0.04} & \textbf{1.00/0.00}\\
 & GTSRB & GoogLeNet & \textbf{1.00/0.00} & 0.98/0.00 & 0.96/0.04 & \textbf{1.00/0.02} & \textbf{0.76/0.00} & 0.74/0.04\\
 & ImageNet-R & ResNet-50 & \textbf{1.00/0.00} & \textbf{1.00/0.00} & \textbf{1.00/0.00} & \textbf{1.00/0.00} & \textbf{1.00/0.00} & \textbf{1.00/0.00} \\
 & MNIST & CNN-7 & \textbf{1.00/0.00} & \textbf{1.00/0.00} & \textbf{0.97/0.02} & \textbf{1.00/0.00} & \textbf{0.97/0.02} & \textbf{1.00/0.00}\\
\bottomrule
\end{tabular}
\end{table*}

\subsection{Out-of-Setting Detection Baselines}
\label{sec:outofsetting}

Lastly, we compare DeBackdoor to a range of state-of-the-art techniques that do not operate in our pre-deployment, data-limited, single-instance, and black-box setting. ABS and Neural Cleanse (NC) are introduced in Section~\ref{sec:TDC}.  ANP~\cite{wu2021adversarial} operates similarly to ABS, applying adversarial perturbations to model neurons in order to identify abnormal patterns in the model's output that are indicative of a backdoor. DF-TND~\cite{wang2020datafree} operates similarly to NC, but generalizes to cases where no clean validation images are available. STRIP~\cite{gao2019strip} is a post-deployment technique that detects triggers in the input by identifying patterns which are abnormally robust to perturbations. FreeEagle~\cite{fu2023eagle} generates dummy intermediate representations for each class through gradient descent-based optimization and uses anomaly metrics computed from these representations to identify trigger-induced behaviors.

DeBackdoor outperforms these baselines in most backdoor settings (Table~\ref{tab:white-box}). These backdoors include trigger types that are effectively covered by the generic templates included in DeBackdoor, and target label functions that are supported by DeBackdoor (All2One/One2One).

We also evaluate DeBackdoor against a semantic attack that employs a natural trigger~\cite{fu2023eagle}. In this attack, the attacker selects a natural semantic feature that can be found in clean images of a victim class as the trigger (e.g. a green meadow in the image of a ram). Then, the attacker injects the backdoor into the model such that any image of a ram in a green meadow is misclassified to a specific target class, while all other images of rams are correctly classified. DeBackdoor outperforms the baselines by a large margin.

\subsection{Ablation Study}
\label{sec:ablation}

All experiments running our detection technique were conducted using an NVIDIA Tesla V100 GPU and 32GB of RAM. We analyze the effects of the choices of hyper-parameters of our technique on a subset of the TDC dataset: clean image batch size $b$, number of optimization steps $s$, and score smoothing parameter $\lambda$. 

\textbf{Batch size} refers to the number of clean validation images used to compute the cASR. We see that time and AUROC both increase as $b$ grows (Figure~\ref{fig:hyper-parameters}-a). While larger batches take a longer time for inference, increasing $b$ makes cASR a closer estimate of ASR. For our approach, a batch size of $32$ is sufficient.

\textbf{Number of steps} counts the iterations of Simulated Annealing. Figure~\ref{fig:hyper-parameters}-b shows the effect of the number of steps on AUROC. In our experiments, 10000 Simulated Annealing steps have been enough. However, for harder problems, one can potentially use more steps, as the computation takes only a few minutes for a model in our experiments.

$\boldsymbol{\lambda}$ \textbf{parameter} trades off smoothing and similarity when approximating ASR using cASR. We vary $\lambda$ and observe the best performance when $\lambda=0.6$ (Figure~\ref{fig:hyper-parameters}-c). Larger values of $\lambda$ cause discreteness (Equation~\ref{eq:cASR}) while smaller values of $\lambda$ yield inaccurate approximation of ASR. To search for $\lambda$ and batch size, we deliberately reduced the number of Simulated Annealing steps to avoid AUROC $=100\%$, so that improvements become visible.

\begin{figure}[t]
  \centering
  \begin{minipage}[htp]{0.23\textwidth}
    \centering
    \includegraphics[width=1\textwidth]{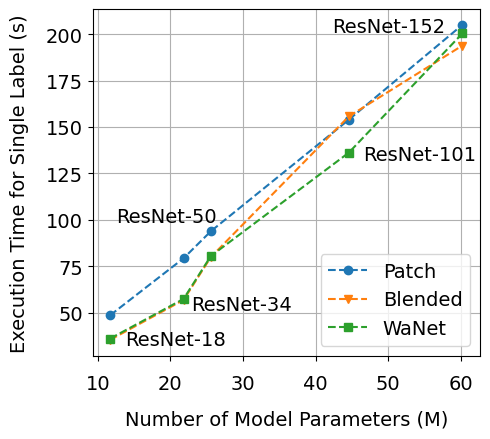}
    (a) ResNets~\cite{he2016deep}
  \end{minipage}
      \centering
  \begin{minipage}[htp]{0.23\textwidth}
    \centering
    \includegraphics[width=1\textwidth]{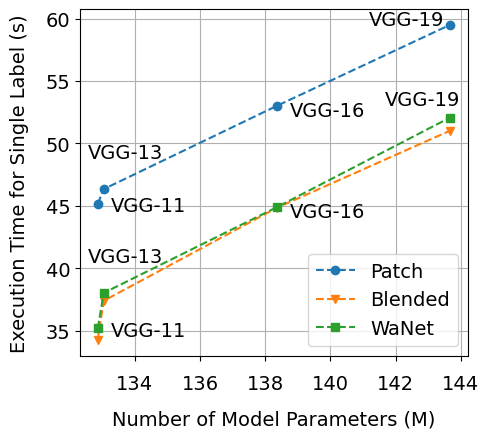}
    (b) VGGs~\cite{simonyan2015vgg}
  \end{minipage}
  \caption{A comparison of detection runtime per class label of different backdoor attacks across different model architectures and varying model sizes.}
  \label{fig:runtime}
\end{figure}

\textbf{Computational cost} of our detection technique is dominated by the runtime of querying the model (Table~\ref{tab:time-comparison}). The runtime is about a few minutes per category using a V100 GPU. Simple architectures (lightweight CNNs) are faster than larger ones (vision transformers). The time also depends on the complexity of the search space as simpler search spaces require fewer SA steps. Figure~\ref{fig:runtime} shows that detection runtime grows linearly, and at different rates (depending on the architecture), as the number of model parameters increases. The search across multiple target labels is straightforward to parallelize.

\begin{table}
  \small
  \centering
  \caption{A comparison of detection runtime per class label for different backdoor attacks, models, and image datasets in the TDC. When 10000 Simulated Annealing steps are used, it takes about or less than a minute on a single V100 GPU.}
  \begin{tabular}{lccc}
    \toprule
    \multirow{2}{*}{Attack} & \multicolumn{3}{c}{Label Inspection Time (s)} \\ \cline{2-4}
           & Wide-ResNet & Vision Transformer & CNN\\
           & (CIFAR-10) & (GTSRB) & (MNIST)\\
    \midrule
    Patch & 66.44 & 60.28 & 16.33 \\
    Blended & 52.62 & 49.32 & 13.60 \\
    WaNet & 17.57 & 14.66 & 5.66 \\
    LIRA  & 8.65 & 4.59 & 1.69 \\
    \bottomrule
  \end{tabular}
  \label{tab:time-comparison}
\end{table}

\section{Discussion}
\label{sec:discussion}

\textbf{Future attacks}. DeBackdoor behaves similarly to Antivirus software. Given a model, DeBackdoor scans it to find potential backdoors given a known list of broad attack templates. Each template defines a search space that DeBackdoor searches through. Here, future-proofness depends on the database of templates and we may miss a zero-day attack unless the defender introduces an appropriate template. We highlight that DeBackdoor is easily adaptable to incorporate new templates to cover novel attack types as they arise. This being said, our experiments show that if an unseen new attack behaves similarly to an existing template, the new attack can still be detected. For example, even though MMS-BD~\cite{kwon2022multi} uses location-specific triggers and c-BAN~\cite{salem2022dynamic} is a dynamic attack, DeBackdoor can reliably detect both of these newer attacks via the generic patch template.

\textbf{Defining a new attack}. We argue that whether a behavior is considered an attack or not, is a matter of definition. This definition should be expressed in some form. Defense techniques define a backdoor attack using either (1) a set of backdoored models, or (2) a set of triggered inputs, or (3) some broad knowledge about the behavior of the attack. We argue that in the absence of the first two sources, we must rely on the third. 

\textbf{Detection of model watermarks.} Some model watermarking techniques rely on embedding triggers into clean samples, causing the model to classify the sample to a verification output. In this case, DeBackdoor may detect this trigger as it behaves identically to a backdoor. Ultimately, it is up to the defender to decide whether this trigger compromises the model or serves verification purposes.

\section{Related Work}
\label{sec:related-work}

\subsection{Backdoor Attacks}
\label{sec:related-work-backdoor-attacks}

A backdoor attack operates by injecting a hidden backdoor into a deep model. When the model is provided an input that contains a trigger, the model misbehaves by misclassifying the input to an attacker-chosen target label. Otherwise, the model's classification of clean inputs is identical to that of a normal model, making backdoor attacks stealthy.

BadNets~\cite{gu2017badnets} proposed the first backdoor attack that uses a patch as a trigger, causing the backdoored model to misclassify any input with this trigger to an attacker-chosen target label $t$ (sample-agnostic). The backdoor is injected into the model by poisoning a subset of the training data samples by adding the trigger to the input and changing the label to $t$. The model is then trained using this data to learn the trigger as a strong feature of the target class.

Some works focused on making this attack more stealthy by developing new triggers. For example, the blended attack~\cite{chen2017targeted} proposes using a trigger that is blended into the background of the input and thus more difficult to distinguish. Refool~\cite{liu2020reflect} proposes planting triggers into inputs that appear as natural reflections, bypassing visual inspection.

Other works focused on making backdoor attacks more practical by developing triggers that are effective in real-world scenarios. Physical and semantic backdoors~\cite{li2021rethinking, xue2021facial, fu2023eagle, wenger2021physical} propose using physical objects (e.g. pink sunglasses) and semantic features (e.g. green meadows) as triggers that can be found in the real world. Compression-resistant backdoors~\cite{xue2022resistant} design triggers that are resilient to changes induced during the compression process.

Recently, dynamic attacks were proposed as a powerful alternative to static (sample-agnostic) attacks~\cite{salem2022dynamic}. Dynamic attacks use a trigger that is sample-specific. For example, WaNet~\cite{nguyen2021wanet} proposes using a warping effect as a trigger, which produces different changes for each input. LIRA~\cite{doan2021lira} trains an autoencoder to generate noise that is distinct for each input and invisible to human inspection. These attacks bypass detection techniques that assume a single trigger is used across all inputs. In addition to sample-specific triggers, some works proposed making the target label dependent on the input~\cite{kwon2022multi}. 

Since poisoned samples can be detected in the training dataset by flagging inputs with incorrect labels, clean-label attacks were proposed as an alternative that does not require the attacker to modify labels~\cite{barni2019cleanlabel, turner2019labelconsistent, saha2020hidden, zhao2022classoriented}. We defer the review of backdoor attacks in other domains to appendix~\ref{appendix:C}.

\subsection{Backdoor Defenses}

A wide range of methods have been proposed to defend against backdoor attacks. These methods operate in different defense settings and take diverse approaches to address the threat of backdoor attacks.

Some defense techniques are deployed alongside the model and defend the model by inspecting and either sanitizing or filtering out any inputs that might contain a backdoor trigger~\cite{guo2023scale, veldanda2021nnoculation, doan2020februus}. For example, STRIP~\cite{gao2019strip} applies perturbations to incoming inputs, observes the randomness of predicted classes, and identifies low entropy in predictions as an indicator of the presence of a backdoor trigger.

Other defense techniques are pre-emptive, defending the model by analyzing and isolating the subset of the training data that is poisoned~\cite{qi2023proactive, tran2018spectral, zhang2021topological, pan2023asset, udeshi2022model}. For example, AC~\cite{chen2018activation} clusters the training dataset using the model activations for different inputs. The cluster of poisoned samples is then detected and can be excluded from model training.

A few defense techniques build classifiers that classify models as clean or backdoored. These techniques assume that a dataset of clean and backdoored models is available (e.g. setting of TDC~\cite{neurips2022tdc}). Otherwise, these techniques train datasets of clean and backdoored models~\cite{xu2021mntd, kolouri2020litmus}.

Some defense techniques do not rely on any triggered inputs or training data, detecting backdoors by directly inspecting the internals of the model~\cite{liu2019abs, gao2019cost, wang2020datafree, zhang2021topological}. For example, FreeEagle~\cite{fu2023eagle} uses dummy inputs and records the intermediate representations of the model to detect anomalous activations that indicate the model contains a backdoor.

A family of defense techniques attempts to reverse-engineer the trigger~\cite{wang2019cleanse, tao2022pixel, shen2021karm, dong2021black, guo2021aeva, wang2020datafree}. For each potential target class, a candidate trigger is synthesized by solving the backdoor optimization problem (i.e. finding a small perturbation that causes the model to misclassify inputs to the target class). Once a set of candidate triggers is synthesized, various anomaly detection methods are applied to identify whether a true backdoor trigger exists.

Beyond detecting backdoor attacks, a range of defense techniques have been proposed that aim to remove a backdoor from the infected model~\cite{huang2024ubainf, liu2024mudjacking, liu2018finepruning, wu2021adversarial, zhao2020bridging}. Some of these techniques propose removing backdoors by fine-tuning the model on clean data. The other set of these techniques removes the backdoor by pruning backdoored neurons and thus preventing the backdoor from being activated.

\section{Conclusion}
\label{sec:conclusion}

In this work, we consider a realistic defense scenario against backdoor attacks. This scenario is defined by a few characteristics that we refer to as the pre-deployment, data-limited, single-instance, and black-box constraints.

Given a query model, our framework generates candidate triggers by optimizing a novel objective function. This yields explainable results that can be used to detect whether the model has a backdoor, as well as the trigger, target labels, and the success rate of the attack. We generate triggers using Simulated Annealing, which is appropriately designed for non-convex optimization problems.

Most prior works do not operate in our scenario. However, for evaluation and documentation purposes, we compare our framework against a range of detection techniques that are not restricted to these constraints. We evaluate our framework against eight attack techniques, five architectures, and five datasets. Our model demonstrate superior performance across these diverse settings.

\newpage

\section*{Ethics Considerations and Compliance with the Open Science Policy}
 
\subsection*{Ethics Considerations}
 
In developing \Z, no human subjects were involved, so there are no ethical concerns related to data privacy. However, the ethical implications of this work are significant, given its focus on providing a novel method for detecting backdoor attacks in deep learning models that may be used in safety-critical applications. This work intends to provide an additional method for verifying the safety and security of this technology before its deployment. Additionally, by highlighting these threats, this study raises awareness within the research and practice communities and expedites the development of robust defensive mechanisms. Our work positively contributes to secure and trustworthy AI, ensuring deep models are safer for all applications and user communities.
 
\subsection*{Compliance with the Open Science Policy}
 
We provide sufficient details regarding configurations, datasets, baselines, metrics, and deep models used in our study and experiments. These details greatly support and enhance the reproducibility and replicability of our scientific findings. To further facilitate the artifact evaluation in terms of availability, functionality, and reproducibility of our findings, the implementation of \Z\ is publicly available at \href{https://zenodo.org/records/14738587}{https://zenodo.org/records/14738587}. We have included detailed instructions for installation and execution, alongside examples of models backdoored using the attacks discussed in this study. Due to the large size of the complete datasets, they will be available upon request.

\bibliographystyle{plain}
\bibliography{biblio}

\begin{appendices}

\section{Implementation Details of Backdoor Attacks}
\label{appendix:A}
For patch attacks, the pattern matrix $p$ is randomly sampled from an independent Bernoulli $0/1$ distribution. The mask matrix $m$ is distinct for each model, so each trigger has a different location and size. For blended attacks, the pattern matrix $p$ is randomly sampled from an independent Uniform $[0,1]$ distribution. For filter attacks, we adopt the negative color filter for 1-channel images (MNIST) and the vintage-photography-style filter for 3-channel images (CIFAR-10, GTSRB, ImageNet)~\cite{fu2023eagle}. These attacks are either All2One or One2One, with a random choice of $t$ and $s$. We use stealthier adaptations of these attacks that fine-tune models from the starting parameters of clean models. Furthermore, regularization with multiple similarity losses ensures that clean and backdoored models appear statistically indistinguishable. We consider diverse triggers (pattern, shape, size, target label).

For WaNet attacks, the $k\times k\times2$ control-grid is randomly sampled from an independent Uniform $[-1,1]$ distribution. $s$ is randomly sampled from an independent Uniform $[0.25,0.75]$ distribution and $k$ is randomly sampled from the set $\{3,4,5,6\}$, as values outside of these ranges either make the trigger visually distinguishable or ineffective. For LIRA attacks, the transformation function $T_{\xi^{*}}$ is learned using a simple autoencoder architecture. These attacks are either All2One with random choice of $t$, or All2All with one-shift strategy. 

MMS-BD targets $2\leq m\leq10$ models and injects each with a location-specific patch trigger. c-BaN injects a single model with $2\leq n\leq10$ sample-specific patch triggers generated from a uniform distributions and using random locations. These attacks are All2One with random choice of $t$.

For semantic attacks, the trigger is a green meadow, a natural semantic feature of images in the ``ram'' class. These attacks are One2One, with random target class $t$.

\section{Texture Similarity of Original and Synthesized Triggers}
\label{appendix:B}

In Section~\ref{sec:TDC}, we evaluated the performance of DeBackdor on the trigger synthesis task by measuring the intersection over union (IoU) between the mask of the original trigger and the mask of the trigger synthesized by DeBackdoor. In addition to measuring the similarity of triggers masks, we also measure the similarity between the textures of original and synthesized triggers. To measure the similarity between two textures, we extract a 4096 dimensional feature vector from each trigger. We used bag of words features because they are agnostic to texture size. Finally, we calculate the cosine similarity (CS) between the two textures.

We calculate CS between our predicted texture and the ground-truth texture, and between a random texture and the ground-truth texture. For each of the 4 datasets in TDC, we calculated this similarity analysis on 125 backdoored models and reported the results in table~\ref{tab:texture-similarity}. Our predicted textures are more similar to the ground truth than a random texture. However, they are very different from the ground truth texture. It is notable that even though our predicted textures are different from the ground truth, they are usually more effective (i.e. higher ASR). This highlights that a byproduct of the backdooring process is the injection of not only the attacker-chosen trigger, but a range of triggers of varying sizes and textures.

\begin{table}[t]
\small
\setlength{\tabcolsep}{4pt}
\caption{The similarity score of our synthesized triggers for patch attacks in the TDC. For each of the similarity scores, 125 different triggers on 125 different models are used.}
\label{tab:texture-similarity}
\centering
\begin{tabular}{l l c c}
\toprule
\multirow{2}{*}{Dataset} & Model & \multicolumn{2}{c}{Cosine Similarity (CS)} \\
\cline{3-4}
 & Architecture & Random & DeBackdoor \\
\midrule
CIFAR-10 & Wide ResNet-40-2 & 0.007 & 0.021 \\
CIFAR-100 & Wide ResNet-40-2 & 0.012 & 0.041 \\
GTSRB & ViT & 0.013 & 0.034 \\
MNIST & CNN-5 & 0.008 & 0.842 \\
\bottomrule
\end{tabular}
\end{table}

\section{Further Discussion}
\label{appendix:D}
\textbf{Beyond the vision domain}. We evaluate our detection technique on numerous computer vision models. In this work, we have demonstrated the versatility of our framework by showing that simple modifications to the trigger search space enable the detection of new attacks. Future work can explore extending our framework to encompass various models in other domains such as text, speech, and graphs. In these cases, in addition to modifying the trigger search space, several low-level submodules will also need modification to fit the target domain. For example, the backdoor application submodule will be adapted to apply perturbations on the target domain, and the neighbor generation submodule will be adapted to generate neighbors based on changes in the target domain.

\textbf{Decision boundary for detection}. In realistic scenarios, a defender will inspect a model that they have received from an untrusted third party and receive a single score for this model. While using a hard threshold of $95\%$ is generally effective, we highlight two cases where this threshold will lead to issues. First, when the trigger used in an attack does not have an explicit search space and must be approximated (e.g. LIRA), the scores produced by our technique will drop for both clean and backdoored models. This means that scores of some backdoored models will drop below the $95\%$ cutoff, leading to an increase in false negatives. However, the scores of all backdoored models will still be significantly higher than any clean model. Consequently, when performing approximate detection of attacks, the defender can lower the decision boundary and achieve high detection performance. 

Second, the scores produced by our technique are dependent on the upper bound of the perturbation $\delta_{S}$ that is used as the attacker's trigger. If the bound is very high, we observe that our technique is capable of synthesizing effective triggers with high scores for even clean models. Intuitively, if you allow a trigger to perform significant perturbation to the image (e.g. a patch that covers half of the image), then the trigger can alter the image's features to make it look like it belongs to a different class.

\section{Emerging Threats}
\label{appendix:C}

Deep models are increasingly employed in tasks across different domains. This poses a diverse set of security problems, with backdoor attacks being demonstrated against audio~\cite{chen2024audio}, graph~\cite{zhang2021graphs}, language~\cite{dai2019text}, and video~\cite{hammoud2024video} 
models. Recent backdoor attacks against large language models (LLMs)~\cite{yan2024easyllm, zhang2024instructllm} are of special concern due to the widespread adoption of LLMs in safety-critical tasks such as code generation.

Backdoor attacks have also been implemented in a wide range of deep learning paradigms (e.g. contrastive learning~\cite{li2024contrastive}, reinforcement learning~\cite{wang2021reinforce}, transfer learning).

Federated Learning (FL) is especially vulnerable to backdoor attacks. In FL, untrusted clients participate in training a global model. The client performs a local round of training on their data and shares only the resulting gradients with a server. The server aggregates these gradients from many clients and uses them to update the global model. This setting makes backdoor attacks a pressing concern in FL as any of the clients can inject a backdoor, with both attacks and defenses being demonstrated in practice~\cite{li2024detectfed, lyu2024attackfed, bagdasaryan20federated}.

Overall, backdoor attacks are adaptable to many different settings and will continue to pose a threat to deep models and their users. Given the rapid advancement and adoption of new deep models, research on backdoor attacks and defenses will remain an active research direction in the coming years.

\begin{table}[t]

\caption{An overview of the 20 classes used from ImageNet~\cite{deng2009imagenet} to construct ImageNet-R.}
\label{tab:imagenetr}
\centering
\begin{tabular}{c c c}
\toprule
 Class ID & Class ID & Class \\
 in Imagenet-R & in ImageNet & Description \\
\midrule
0 & n02114367 & grey wolf\\
1 & n02123159 & tiger cat\\
2 & n02342885 & hamster\\
3 & n02412080 & ram\\
4 & n02894605 & breakwater\\
5 & n02895154 & breastplate\\
6 & n02930766 & taxicab\\
7 & n02999410 & chain\\
8 & n03089624 & confectionery store\\
9 & n03125729 & cradle\\
10 & n03141823 & crutch\\
11 & n03201208 & dining table\\
12 & n03240683 & drilling rig\\
13 & n03450230 & gown\\
14 & n03773504 & missile\\
15 & n03787032 & square academic cap\\
16 & n03792782 & mountain bike\\
17 & n03929855 & pickelhaube\\
18 & n03937543 & pill bottle\\
19 & n04162706 & seat belt\\
\bottomrule
\end{tabular}
\end{table}

\end{appendices}

\end{document}